\documentclass[fleqn,usenatbib]{mnras}

\usepackage[T1]{fontenc}
\usepackage{lipsum}  
\DeclareRobustCommand{\VAN}[3]{#2}
\let\VANthebibliography\thebibliography
\def\thebibliography{\DeclareRobustCommand{\VAN}[3]{##3}\VANthebibliography}

\usepackage{graphicx}	
\usepackage{amsmath}	
\usepackage{amssymb}	
\usepackage{acro}
\usepackage{caption}
\usepackage{enumitem}
\usepackage{pifont}
\usepackage{bigints}
\usepackage{ulem}
\usepackage{hyperref}
\hypersetup{
    colorlinks=true,
    linkcolor=blue,
    filecolor=magenta,      
    urlcolor=cyan,
    pdftitle={Gravity can bind the ISM but not the CGM},
    pdfpagemode=FullScreen,
    }

\usepackage[dvipsnames]{xcolor}
\usepackage{arydshln}
\usepackage{booktabs}
\usepackage{adjustbox}

\newcommand{\xmark}{\ding{55}}
\newcommand{\rfchange}[1]{{#1}}
\newcommand{\adchange}[1]{{#1}} 

\newcommand{\clicks}[2]{\href{#1}{\colorlet{temp}{.}\color{blue}{\color{temp}#2}\color{temp}}}
\graphicspath{ {./images/model1} }
\graphicspath{ {./images/model2} }
\defcitealias{RB2006MNRAS}{RB06}

\title[Gravity can bind the ISM but not the CGM]{Ram pressure stripping in clusters: Gravity can bind the ISM but not the CGM}

\author[Ghosh et al.]{
Ritali Ghosh$^{1}$\thanks{E-mail: ritalighosh@iisc.ac.in},
Alankar Dutta$^{2}$ 
and
Prateek Sharma$^{1,2}$
\\
$^{1}$Joint Astronomy \& Astrophysics Programme, Indian Institute of Science, Bangalore 560012, India\\
$^{2}$Department of Physics, Indian Institute of Science, Bangalore 560012, India
}

\date{Accepted XXX. Received YYY; in original form ZZZ}

\pubyear{0000}

\usepackage{newtxtext,newtxmath}
\usepackage{listings}
\usepackage{threeparttable}
\begin{document}
\label{firstpage}
\pagerange{\pageref{firstpage}--\pageref{lastpage}}
\maketitle

\begin{abstract}
We explore the survival of a galaxy's circumgalactic medium (CGM) as it experiences ram pressure stripping (RPS) moving through the intracluster medium (ICM). For a satellite galaxy, the CGM is often assumed to be entirely stripped/evaporated, an assumption that may not \adchange{always} be justified. We carry out 3D-hydrodynamic simulations 
of the \rfchange{interstellar and circumgalactic media (ISM+CGM)} 
of a galaxy 
like JO201 moving through the ICM. 
The CGM \adchange{can} survive long at cluster outskirts ($\gtrsim2 \rm \ Gyr$) but at smaller cluster-centric distances, 90\% of the CGM mass is lost within $\sim 500$ Myr. 
The gravitational restoring force on the CGM is mostly negligible and the CGM-ICM interaction is analogous to \textit{`cloud-wind interaction'}. The CGM stripping timescale \adchange{does not depend on the ram pressure but on the CGM to ICM density contrast $\chi$}. Two distinct regimes emerge for CGM stripping: the $\chi >1$ regime, which is the well-known \textit{`cloud crushing'} problem, and the $\chi <1$ regime, which we refer to as the (relatively unexplored) \textit{`bubble drag'} problem.
\adchange{The first pericentric passage near the cluster core can rapidly --  over a crossing time $t_{\rm drag} \sim R/v_{\rm rel}$ -- strip the CGM in the \textit{bubble drag} regime.} 
The \rfchange{ISM} stripping criterion \adchange{unlike the CGM criterion,} still depends on the ram pressure $\rho_{\rm ICM} v_{\rm rel}^2$. 
The stripped tails of satellites contain contributions from both the disk and the CGM. 
\adchange{The 
X-ray plume in M89 in the Virgo cluster and a 
lack of it 
in the nearby M90 
might be attributed to their orbital histories. M90 has likely undergone 
stripping in the bubble drag regime due to a pericentric passage close to the 
cluster center. 
}
\end{abstract}




\section{Introduction}
    A galaxy's environment plays a significant role in shaping the distribution of matter around it. While internal processes regulate the gas content of a galaxy by continuously mixing pristine intergalactic matter and metal-rich ejecta from stars through inflows and outflows (\citealt{Sarkar2015-10.1093/mnras/stu2760, Fielding2016-10.1093/mnras/stw3326}), the external dense environment around it (the intragroup/intracluster medium) can dramatically affect this evolution (\citealt{Dressler1980ApJ, Poggianti1999ApJ...518..576P}). 
    Surveys of local and distant clusters suggest that the fraction of red, quiescent, and gas-deficient galaxies is elevated in the clusters compared to their field counterparts (\citealt{Boselli2006, Peng_2010, Boselli2014}). 
    \adchange{This environmental effect becomes pronounced as a satellite galaxy moves closer and closer to the cluster's potential well.
    Observations by \citealt{ButcherOemler1984ApJ} and subsequent works on the environmental effects in the SDSS, GOGREEN, zCOSMOS, and other deep surveys show a suppression in the star-formation rate with decreasing cluster-centric distances (\citealt{ Mahajan2012, Burg2020A&A...638A.112V}). }
    The external environment, therefore, modulates a galaxy's morphology, \adchange{color and} its star-formation history.
    
    While moving through the intracluster medium (ICM hereafter), both hydrodynamic and gravitational interactions affect a galaxy, bringing out broadly two changes:
    \begin{enumerate}[wide, topsep=0.25ex, itemsep=0.25ex]
        \item \textit{Gravitational or tidal interactions} with other cluster members (\citealt{Merritt1983ApJ...264...24M}), mergers, and harassment (high-speed encounters between galaxies) (\citealt{Moore1996, Mahajan2012}) \rfchange{perturb} both the gaseous and stellar distribution of a galaxy. Since merger cross-sections are low in clusters (\citealt{Makino_1997}), and tidal interactions are short-lived to produce severe disturbances, these interactions are not sufficient to explain the above-mentioned gradient in the star-formation rate (\citealt{Boselli2006}).
        
        \item \textit{Hydrodynamic interactions} with the hot ICM can create a disturbed gaseous distribution in the wake of the galaxy's orbital path. \adchange{Specifically, the moving galaxy experiences ram pressure from the surrounding ICM that strips off its gaseous components} (\citealt{GunnGott1972ApJ}) without perturbing the stellar distribution. This can result in long tails of ionized gas trailing behind, with the most spectacular cases referred to as `\textit{jellyfish galaxies}'. 
        The observed tails are multiphase - the ionized ($\rm H\alpha$: \citealt{Fumagalli2014,Poggianti_2019};
        UV: \citealt{Smith2010, Gullieuszik_2023, George2023MNRAS};
        X-ray: \citealt{Sun_2010}), \adchange{the} neutral (HI: \citealt{Cayatte1994AJ....107.1003C, Poggianti2016AJ....151...78P}),
        and even \adchange{the} molecular (CO: \citealt{Jachym_2019, Moretti2023ApJ}) phases co-exist (see \citealt{Sun2022} and references therein). Ram pressure stripping (RPS hereafter) is the most plausible \adchange{generic} mechanism to remove the galaxy's cold gas reservoir and quench it as it moves through the ICM.
    \end{enumerate}
    
    Numerical simulations in the past few decades have intensively examined the stripping of dense interstellar medium (ISM hereafter) \adchange{of a galaxy falling through} the hot ICM (\citealt{Roediger2006,  Roediger2007, Jachym2009A&A...500..693J, Tonnesen&Bryan2010ApJ, Tonnesen2021ApJ...911...68T, Lee2022ApJ...928..144L, Choi2022ApJ...936..133C}). These works have been able to successfully explain some of the observed signatures like the truncation radius of jellyfish galaxies (radius where gravitational restoring pressure of the galaxy equals the ram pressure - the classic \citealt{GunnGott1972ApJ} criterion). Stripping of the \adchange{ISM} is found to be dependent on its density structure. In line with observations, \adchange{the} diffuse \adchange{ISM} is stripped easily compared to the dense molecular clouds (\citealt{Kenney1989ApJ...344..171K, Fumagalli_2009, Abramson_2014, Moretti2020ApJ...889....9M, Bacchini2023}). Given the vast range of scales involved in the problem, most idealized simulations often focus on only the ISM-ICM interaction. 
    
    Galaxies harbor not only a dense \adchange{gaseous disk constituting the ISM}, but also a massive, extended, and diffuse medium around it - the circumgalactic medium (CGM hereafter). 
    \adchange{This diffuse CGM, in fact,} contains the majority of baryons associated with the host halo of the galaxy (\citealt{Tumlinson/annurev-astro-091916-055240}, \citealt{Bregman_2018}). The CGM brings in pristine gas from the intergalactic medium \rfchange{(IGM henceforth)} and feeds and recycles material to \adchange{and from} the galactic disk. Losing such a huge reservoir over cosmological timescales would result in a restrained supply of cold gas to the central disk needed to sustain star formation. Complete stripping of the CGM can lead to \textit{strangulation}, whereby the galaxy quenches as the supply of cold gas to the disk is 
    \adchange{removed} (\citealt{Larson1980ApJ...237..692L, Woo2015MNRAS.448..237W}).
    
    \adchange{The significance of considering a hot halo/CGM of a satellite galaxy in cluster environments is also highlighted by the semi-analytic galaxy formation models.} The galaxy formation models in the 1990s, \adchange{presumed the hot halo(/CGM) of a satellite to be completely stripped when it enters the virial radius of the host cluster.} This resulted in the fraction of quiescent galaxies to be much higher than the observed fraction (\citealt{Kauffmann1993, Cole1994, Baldry2006, Weinmann2006}). Subsequent semi-analytic models suggest that retaining a fraction of hot gas/CGM \adchange{of a satellite on its first infall}
    (\citealt{Font2008, Guo2011, Stevens2016}) is crucial for concordance with the observed quiescent fraction in clusters (\citealt{Boselli2006, Boselli2014}). 
    \adchange{Recent cosmological zoom-in simulations from {\tt TNG-Cluster} show that some massive cluster satellites can retain their CGM, potentially shielding the satellite ISM from being directly stripped (\citealt{Rohr2024}). Accurately modeling environmental quenching mechanisms, therefore, requires properly treating the diffuse CGM around satellite galaxies, not just the dense ISM}.
    
    \adchange{Idealized simulations modeling the stripping of the ISM and the CGM together (ISM+CGM, henceforth) can address numerous additional questions posed by observations and cosmological simulations.}
    For instance, observed stellar metallicities of nearby SDSS galaxies suggest that cluster galaxies quench over a period of 4 Gyr \citep{Peng2015}.
    It is unclear if the galactic disk can continue forming stars for such a long time 
    \adchange{without replenishment from the circumgalactic reservoir.}
    Additionally, the diffuse CGM is thought to be pre-processed in the IGM filaments 
    and stripped even when the galaxy is far from the host cluster ($3-4\ r_{\rm 200}$) (\citealt{Mohammadreza2019}). However, the physical process that drive\adchange{s} the interaction of ISM+CGM with the ICM or the IGM is not well understood. 
    Moreover, ISM+CGM stripping is not only crucial \adchange{for satellites in clusters} but also \adchange{for satellites in} lower mass halos, like our Milky Way. 
    Recent observations of the Magellanic group (Large and Small Magellanic clouds) undergoing RPS in the Milky Way indicate the existence of a (truncated) Magellanic Corona/CGM around the \rfchange{Large Magellanic Cloud (LMC)} \citep{Krishnarao2022Natur.609..915K}.
    Understanding the co-evolution of ISM and CGM in RPS can therefore shed light on galaxy quenching mechanisms as well as the nature of stripped tails in both cluster and group environments.
    
    \adchange{In this paper, we carry out a systematic study of the RPS of a galaxy, with mass and size comparable to the jellyfish galaxy JO201, as it faces a constant ram pressure from the ICM wind}. 
    We carry out 3D hydrodynamical simulations of RPS (without radiative cooling) in the rest frame of the galaxy. Our idealized simulations resemble a \textit{`wind-tunnel}' setup in which the ICM wind properties are \adchange{imposed} all over the domain beyond the virial radius of the galaxy. 
    Within the virial radius, we include both the ISM and the CGM. 
    
    \adchange{To isolate the underlying physical process that determines the removal of the CGM (thereby quenching the galaxy), we carry out an additional set of control simulations that neglect the satellite’s gravitational potential.} 
    The CGM, being diffuse and extended, is loosely bound to the galaxy except for regions close to the disk. \adchange{Our control simulations suggest} that the CGM-ICM interaction is analogous to 
    the idealized \adchange{problem of} cloud-wind interaction. The cloud crushing problem in the context of cold clouds in outflows (\citealt{Klein94, Armillotta2017MNRAS.470..114A, GronkeOh2018MNRAS.480L.111G, Li2020MNRAS.492.1841L, Sparre2020MNRAS.499.4261S, Kanjilal2021MNRAS.501.1143K, Abruzzo2022ApJ...925..199A}) is a subset of such \textit{`cloud-wind interactions'} with cold dense clouds moving in a hot diffuse wind. A wider and somewhat different parameter space is 
    applicable for CGM-ICM interaction. 
    The typical densities of the CGM range from $10^{-5}$ to $10^{-3}\ \rm cm^{-3}$ (for Milky Way and external galaxies), which can be lower than the typical ICM densities- $10^{-3}$ to $10^{-1}$ $\rm cm^{-3}$ (for Virgo-like clusters). \adchange{In this work, we therefore,} explore an alternate regime of the \textit{`cloud-wind interactions'} - the regime of density contrast $\chi<1$; $\chi$ being the density ratio of the CGM 
    to the ICM. 
    We refer to this regime as the `\textit{bubble drag}' regime and elaborate \adchange{upon its physics and relevance to CGM stripping.} 
    
    \adchange{The paper is outlined as follows.} In Section \ref{sec:Framework}, we present 
    a general analytic framework to model a galaxy with three different components (ISM, CGM, and ICM) in hydrodynamic equilibrium. 
    Section \ref{sec:simulationSetup} elaborates \adchange{upon}  the simulation setup. Section \ref{sec:strippingResults} discusses the co-evolution of the ISM and the CGM under \adchange{a constant ram pressure from the hot ICM.
    \adchange{Section \ref{subsec:CloudCrushingComparison} presents the comparison with} \textit{`cloud-wind interaction'} simulations (neglecting the satellite's gravity)}. 
    Section \ref{subsec:TimeScaleBubble} discusses the time scales for the CGM evolution, with an emphasis on the relatively unexplored astrophysical regime of $\chi<1$ 
    (the bubble drag problem). The astrophysical implications and limitations of this work are discussed in Section \ref{sec:discussion}. We conclude in Section \ref{Conclusions} with a summary of the key results.

\section{Galactic ISM+ CGM Model}{\label{sec:Framework}}
\adchange{It is non-trivial to initialize a realistic equilibrium ISM+CGM moving relative to an ICM wind.} In this section, we describe our analytic model for the equilibrium distribution of gas in the ISM+CGM of a galaxy placed within a \rfchange{hot} ICM. We consider the \adchange{galaxy's ISM, CGM and }its host ICM as three isothermal components. For an assumed temperature distribution, our simple yet robust model generates the density distribution for the galaxy with a rotationally supported ISM, surrounded by a 
\adchange{static} CGM 
that smoothly transitions to the ICM beyond the virial radius.

In this section, we first \adchange{highlight} the need for a simple analytic model for the ISM+CGM distribution in idealized RPS studies. \adchange{We} motivate our assumptions and elaborate \adchange{upon} the procedure to analytically solve for the density distribution in Section \ref{subsec:equilibriumSoln}. The details of \adchange{the} assumed gravitational potential and \adchange{the} temperature distribution are mentioned in section \ref{subsec:TempFieldInitialization}. 
Readers 
may skip the details of the analytic modeling and jump straightaway to the final form of the solution \adchange{presented in} Eq. \ref{eq:rhoSolution}, without the loss of continuity.

It is common for the simulations of RPS to ignore the CGM altogether. \adchange{For example}, simulations of ISM-ICM interaction by \citealt{Roediger2006, Tonnesen&Bryan2010ApJ}, prescribe the density, pressure, and rotation velocity in the galactic disk such that it is in pressure equilibrium with a homogeneous ICM (with an assumed ICM density and pressure) around it. Since the ISM pressure is much lower than the ICM pressure, one has the freedom to either choose a lower temperature for ICM wind (compared to observed values) or to have extremely high temperature 
for the disk (see Fig. 5 of \citealt{Roediger2005}). \adchange{
This approach does not provide enough freedom to set the temperatures in the ranges
observed for the ISM and the CGM}. 

Alternatively, one can superimpose simple profiles for the CGM and the ISM. 
However, 
most of these approaches \adchange{($\beta$-profile used by \citealt{Steinhauser2016} or an adiabatic/isothermal profile used by \citealt{Schneider_2018})} 
 produce quasi-equilibrium profiles and are not self-consistent steady solutions of fluid equations. 
Moreover, \adchange{
such approaches (e.g., \citealt{Sarkar2015-10.1093/mnras/stu2760}) 
produce a wide range of temperatures at the ISM-CGM interface, leading to} a cooling-dominated region (if radiative cooling is included) that would collapse in the absence of feedback heating.
Our aim here is, therefore, to avoid these 
pitfalls and utilize 
an equilibrium setup \adchange{with an imposed temperature} to understand the fundamental hydrodynamics of the galaxy-ICM interactions.

    \subsection{Equilibrium solution}\label{subsec:equilibriumSoln}
    We prescribe a temperature field to model the three components - the ISM at $10^4\ \rm K$, the CGM at $2\times 10^6~\rm K$, and the ICM at $10^7~\rm K$.
    With this 
    simple assumption for each component 
    treated isothermally, we obtain a smooth equilibrium density distribution with a rotating ISM and a stationary CGM embedded in a homogeneous ICM. \adchange{The rotation of the gas in the disk is supported by the static gravitational potential provided by the stellar disk and the dark matter halo of the galaxy. Our modeled galaxy has properties motivated by the most spectacular examples of ram pressure stripped galaxies - JO206 and JO201. The different model parameters and their numerical values are listed in Table \ref{tab:Fiducial galaxy params}.}

\begin{figure*}
      \includegraphics[width=0.96\textwidth]{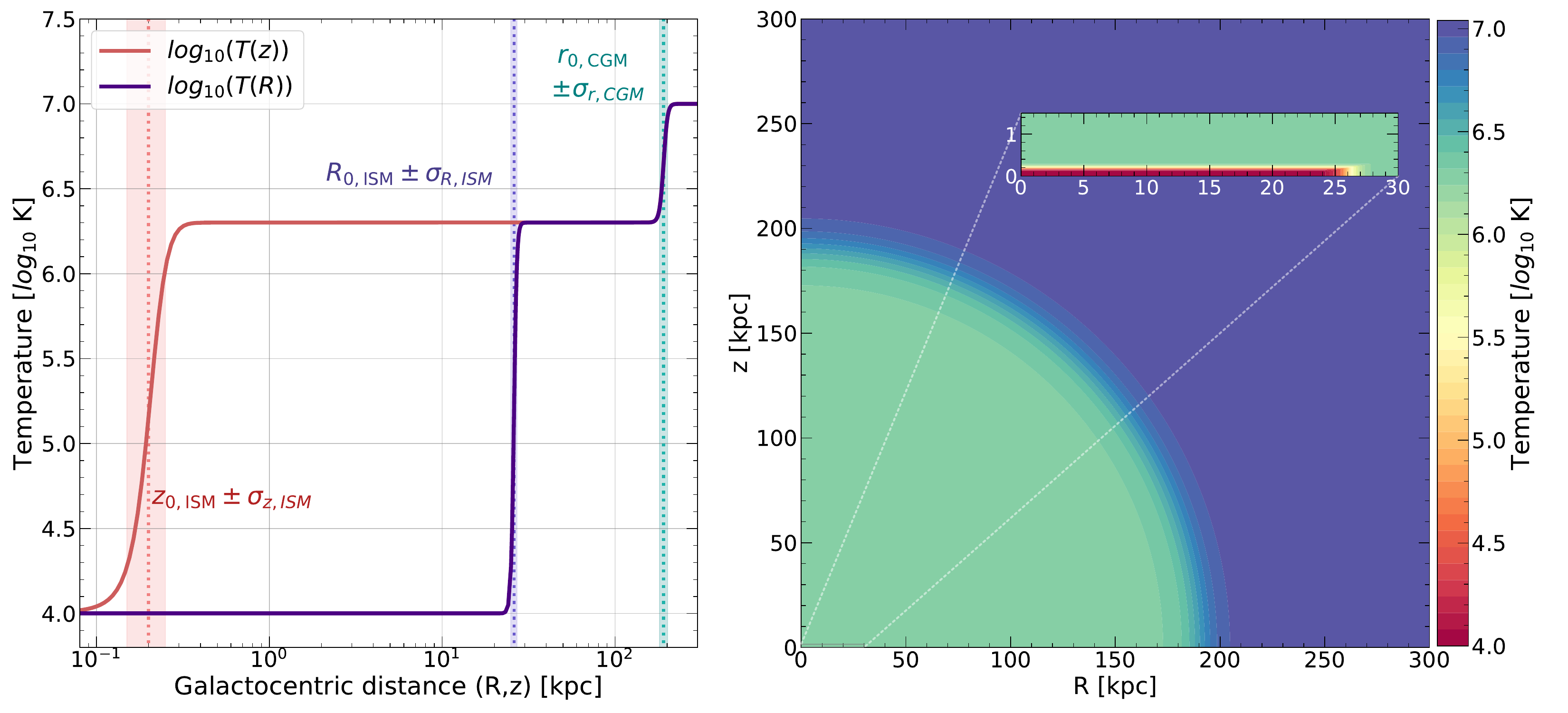}
    \caption{Temperature distribution ansatz for our model galaxy with three isothermal components -- ISM ($10^4$ K) and CGM ($2 \times 10^6$ K) surrounded by a homogeneous ICM ($10^7$ K). \textit{Left Panel}: 1D profiles of temperature ($\log_{\rm 10}T$) as a function of galactocentric distances $R$ and $z$. 
    The \textit{solid red} line denotes the profile along $z$  (at $R=0$ axis) and \textit{solid violet} line is along $R$ (at $z=0$ plane). 
    The \textit{vertical dotted} lines and the shaded regions around them mark the transition zone between the different components
    (\textit{red} and \textit{violet} for ISM-CGM interface along $z$ and $R$ respectively and \textit{cyan} for the CGM-ICM interface; see Section \ref{subsec:TempFieldInitialization}). 
    \textit{Right Panel}: The 2D temperature distribution ($\log_{\rm 10}T$) in $(R, |z|)$ plane. 
    The inset in the right panel shows a zoomed-in region around the galactic disk. 
     }
    \label{fig:temperature profile}
\end{figure*}

We solve for hydrodynamic equilibrium in cylindrical \adchange{polar} coordinates $(R, z)$ centered on the galactic disk 
\adchange{(assuming azimuthal symmetry)}. Our analytic solutions involve integrals that are solved numerically with appropriate boundary conditions\footnote{Details on the numerics related to the initialization of equilibrium profiles can be found in our \href{https://github.com/RitaliG/rps_cgm_hydro}{GitHub repository}.}. 
We use the conventional notation, i.e., $\rho$ for density, $P$ for pressure, $v_{\phi}$ for the azimuthal velocity, and $T$ for the temperature. \adchange{
Now we list the steps to create our initial equilibrium configuration.}
\begin{enumerate}[wide, topsep=0.65ex]
    \item \label{solution: euler} The hydrodynamic equilibrium along the radial and vertical directions mandates,
        \begin{align}
        \hat{R}&:\; -\frac{1}{\rho}\frac{\partial P}{\partial R} - \frac{\partial \Phi}{\partial R} = - \frac{v_{\phi}^2}{R} \label{eq:equilibriumR} \;, \\
        \hat{z}&:\; -\frac{1}{\rho}\frac{\partial P}{\partial z} - \frac{\partial \Phi}{\partial z} = 0\; , \label{eq:EqbZ}
        \end{align} 
    where $\Phi$ is the gravitational potential.
     We begin by assuming an ideal gas equation of state, $P(R,z) =\rho c_s^2(R, z)$, where $c_s=\sqrt{k_BT/\mu m_p}$ is the isothermal sound \rfchange{speed in} 
    the medium at temperature $T$, $\mu$ is the mean molecular weight, $m_p$ is the proton mass and $k_B$ is the Boltzmann constant.
    Eqs. \ref{eq:equilibriumR} \adchange{\&} \ref{eq:EqbZ} can be, respectively, written in the following compact form,
        \begin{align}
            -\frac{\partial \ln \rho}{\partial R} &= \frac{\partial \ln c_s^2}{\partial R} + \frac{1}{c_s^2}\left(\frac{\partial \Phi}{\partial R} - \frac{v_\phi^2}{R}\right) \label{eq:equilibriumR-ln} \;, \\
            -\frac{\partial \ln \rho}{\partial z} &= \frac{\partial \ln c_s^2}{\partial z} + \frac{1}{c_s^2}\frac{\partial \Phi}{\partial z} \; . \label{eq:EqbZ-ln}
        \end{align}
    For a prescribed temperature field $T(R,z)$ (hence $c_s^2$) 
    and $\Phi(R, z)$, these are two equations with two unknowns, namely, $\rho$ and $v_{\phi}$. 
    
    \item\label{step1} We first solve for the equilibrium along the radial direction. Integrating both sides of Eq. \ref{eq:equilibriumR-ln} along $R$, from a reference point $(0,z)$ with density $\rho(0,z)$, we get,
    \[
        \rho(R,z) = \rho(0,z) 
        \frac{c_s^2(0, z)}{c_s^2(R, z)}
        \exp{\left[-\int_0^R \frac{dR'}{c_s^2} \left(\frac{\partial  \Phi}{\partial R'} - \frac{v_\phi^2}{R'}\right)\right]}.
    \]
    \item \label{solution: integrateR} We then solve for the equilibrium along the vertical direction. Integrating both sides of Eq. \ref{eq:EqbZ-ln} along $z$, from a reference point $(R,0)$ with density $\rho(R,0)$, we get,
    \[
        \rho(R,z) = \rho(R,0)
        \frac{c_s^2(R,0)}{c_s^2(R,z)} 
        \exp{\left[-\int_0^z dz'\frac{1}{c_s^2}\frac{\partial \Phi}{\partial z'}\right]}.
    \]
    \adchange{Evaluating} $\rho(R,0)$ from step \ref{step1} 
    and \adchange{substituting it in the above equation gives the following final form of} the density distribution,
        \begin{equation}\label{eq:rhoSolution}
            \begin{aligned}
            \rho(R,z)&={} \rho(0,0) \left(\frac{c_s^2(0,0)}{c_s^2(R,z)}\right)\\
            & \times \exp{\left[
                -\int_0^R dR' \left\{\frac{1}{c_s^2}
                \left( \frac{\partial \Phi}{\partial R'} - \frac{v_\phi^2}{R'}\right)\right\}_{(R', 0)}\right]}\\
            & \times \exp{\left[- \int_0^z dz' \left(\frac{1}{c_s^2}\frac{\partial \Phi}{\partial z'}\right)_{(R,z')}
                \right]}\; .
            \end{aligned}
        \end{equation}
    
    The parameter $\rho(0,0)$ is chosen such that the \rfchange{number density} at the center is \textbf{($\sim 80\ \rm cm^{-3}$)}, which gives a total \rfchange{ISM} mass $M_{\rm ISM}\sim 6.7\times 10^9 M_{\odot}$ and the CGM mass $M_{\rm CGM}\sim 5.4\times 10^{10} M_{\odot}$.
    \footnote{The first exponential term in Eq. \ref{eq:rhoSolution} is approximately always unity, because $v_\phi^2$ is close to $R\partial \Phi/\partial R$ in the disk, as discussed in \ref{vphi-discussion}. It is the second exponential term with integral along $z$ that has a stronger influence on the value of $\rho (R,z)$. To obtain realistic CGM and ICM densities, one has to choose the temperature values carefully, as the integrand is inversely dependent on temperature (through $c_s^2$).}
    The first exponential term in Eq. \ref{eq:rhoSolution} requires the knowledge of gas velocity in the disk plane ($v_{\phi} (R',0)$), which \adchange{needs to be specified.} 

    \item \label{vphi-discussion} 
    The choice of \adchange{the azimuthal velocity} $v_{\phi} (R, z)$ to be plugged in Eq. \ref{eq:rhoSolution} is constrained, 
    since \adchange{Eqs. \ref{eq:equilibriumR-ln} \& \ref{eq:EqbZ-ln}} are \adchange{two }coupled equations \adchange{with} two unknowns ($\rho$ and $v_{\phi}$). 
    Taking the derivative of Eq. \ref{eq:equilibriumR-ln} with $z$ and Eq. \ref{eq:EqbZ-ln} with $R$, subtracting the former from the latter, and integrating along $z$, we get,
    \begin{equation}\label{eq:RotVel2}
        \begin{aligned}
        v_\phi^2(R,z)={}& R\;c_s^2(R, z) \int_{-\infty}^{z} dz'
                          \left[\frac{\partial \Phi}{\partial R} \frac{\partial}{\partial z'}
                          \left( \frac{1}{c_s^2}\right)\right.\\
                        & \; -\left.\frac{\partial \Phi}{\partial z'} \frac{\partial}{\partial R} \left(\frac{1}{c_s^2}\right)\right]_{(R, z')} \; ,
        \end{aligned}
    \end{equation}
    with the assumption that $v_{\phi} (R, -\infty)=0$ (as the CGM and ICM are non-rotating). 
    
    Note that after choosing the rotational velocity at a large distance to vanish, the velocity in the disk plane $v_{\phi}(R,0)$ can not be prescribed independently. Similarly, if one specifies \adchange{the azimuthal velocity in the disk plane} $v_{\phi}(R,0)$, then there would be, in general, a non-zero velocity field beyond the ISM. 
    Eq. \ref{eq:RotVel2} is integrated numerically to obtain the steady-state velocity field for our model.
    The choice of the gravitational potential and temperature distribution used in Eq. \ref{eq:RotVel2} is described 
    in the next section. 

    \item \label{numeics} Substituting 
    $v_{\phi}$ 
    \adchange{(}after solving Eq. \ref{eq:RotVel2}\adchange{) in Eq. \ref{eq:rhoSolution}, 
    and integrating the expression numerically, the density $\rho$ is evaluated at} all $(R, z)$. This, combined with the \adchange{prescribed} temperature \adchange{distribution}, gives the pressure $P(R,z)=\rho (R,z) c_s^2(R, z)$.
\end{enumerate}

    \subsection{The three-phase model parameters}\label{subsec:TempFieldInitialization}
    We prescribe a smooth transition in the temperature from the ISM ($10^4\ \rm K$) to the CGM ($2\times 10^6\ \rm K$) through a \textit{tanh} function, 
    \begin{equation}
        \begin{aligned}\label{eq:temp_dc}
        \Tilde{T}_R &=  \Tilde{T}_{\rm ISM} + 
                       ( \Tilde{T}_{\rm CGM}-\Tilde{T}_{\rm ISM})\frac{1}{2}
                       \left[ 
                            1+\tanh \left( \frac{|z| - z_{0, \rm ISM}}{\sigma_{z,\rm ISM}} \right)
                       \right]\\
        \Tilde{T}_z &=  \Tilde{T}_{\rm ISM} + 
                       ( \Tilde{T}_{\rm CGM}-\Tilde{T}_{\rm ISM})\frac{1}{2}
                       \left[ 
                            1+\tanh \left( \frac{R - R_{0, \rm ISM}}{\sigma_{R,\rm ISM}} \right)
                       \right],\\
        \Tilde{T}'(R,z) &= \frac{1}{2}\left(
                        \Tilde{T}_R +  \Tilde{T}_z + | \Tilde{T}_R- \Tilde{T}_z| 
                        \right),
        \end{aligned}
    \end{equation}
    where  $\Tilde{T}=\log_{\rm 10}(T)$, $\sigma_{R, \rm ISM}$ and $\sigma_{ z, \rm ISM}$ determine the steepness of transition in temperature between the ISM and CGM along $R$ and $z$ respectively.
    This distribution is then used to set the overall temperature field with the ICM values beyond the CGM radius $r_{\rm 0, CGM}$ (chosen to be $200$ kpc for our model galaxy) as in,
    \begin{equation}\label{eq:temp_all}
        \Tilde{T}(R,z) = \Tilde{T}'(R,z)
                        \left[ 
                        1+\left(\frac{\Tilde{T}_{\rm ICM}}{\Tilde{T}_{\rm CGM}}-1\right)\frac{1}{2}
                        \left(1+\text{tanh}\left( \frac{r-r_{0, \rm CGM}}{\sigma_{\rm r,CGM}}  \right)\right)
                        \right] \; ,
    \end{equation}
    where $r=\sqrt{R^2+z^2}$ and 
    $\sigma_{\rm r, CGM}$ quantifies the steepness of transition in temperature between the CGM and ICM. 
    The left panel of Figure \ref{fig:temperature profile} shows the temperature profile of Eq. \ref{eq:temp_all} along $R$ and $z$ using the parameters listed in Table \ref{tab:Fiducial galaxy params}, 
    while the right panel shows the temperature field in the $(R,z)$ plane with a zoom-in to show the disk region in the inset.

    The static gravitational potential of the galaxy is set by the dark matter and the stellar disk, $\rm \Phi = \Phi_{\rm DM} + \Phi_{\rm stars}$, where 
        \begin{equation}
            \Phi_{\rm DM}(R, z) = - \frac{G M_{\rm vir}}{f(c)r_s}
                                 \frac{\ln\left(1+\sqrt{R^2+z^2+d_{\rm halo}^2}/r_s\right)}
                                 {\sqrt{R^2+z^2+d_{\rm halo}^2}/r_s}
        \end{equation}
    is the modified NFW potential (\citealt{ModifiedNFW1997ApJ}) for dark matter for a halo with mass $M_{\rm vir}$, scale length $r_s$ and core radius $d_{\rm halo}$. The factor $f(c)= 
    \ln (1+c) - c/(1+c)$, where $c$ is the concentration parameter. 
    
    The stellar potential is set by the Miyamoto-Nagai profile (\citealt{MiyamotoNagai1975PASJ}) with $a_*$ and $b_*$ as the radial and vertical scale heights, respectively,
    \begin{equation}
        \Phi_{\rm stars}(R, z) = - \frac{G M_{*}}{\sqrt{R^2+\left(a_* + \sqrt{z^2+b_*^2}\right)^2}}.
    \end{equation}
    We have not included a bulge component in our model galaxies as we are interested in 
    stripping of the extended CGM and ISM, 
    and it is well known that the contribution of a bulge enhances the restoring gravitational pressure only near the central regions (\citealt{Abadi}). Neglecting the bulge contribution would, therefore, not affect our results much. 
    \begin{figure}
        \includegraphics[width=1\columnwidth]{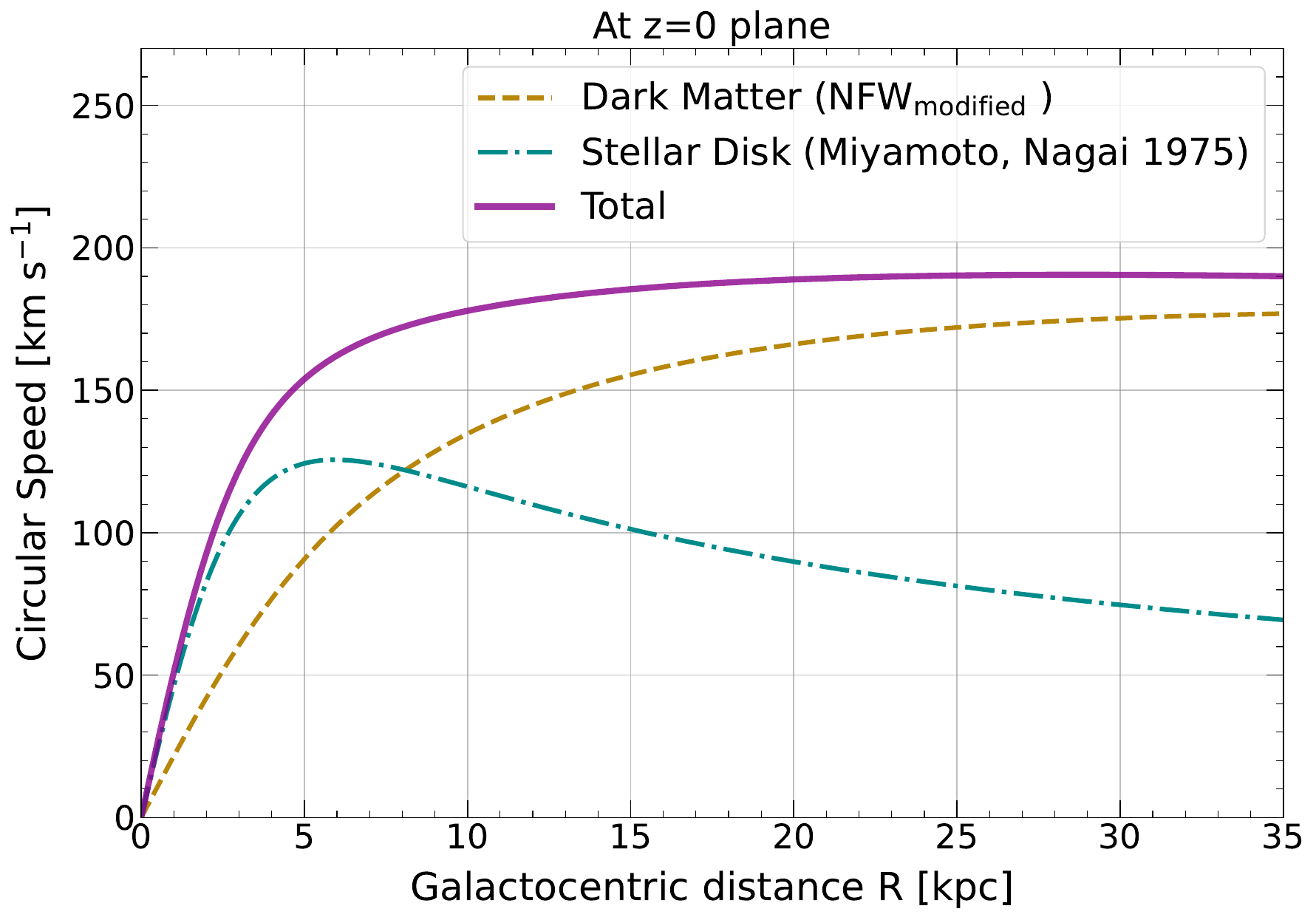}
        \caption{Circular speed in the disk plane ($v_c \equiv R\sqrt{\partial \Phi / \partial R| _{z=0}}]$) as a function of galactocentric distance $R$. Contributions from a modified NFW dark matter halo (\citealt{ModifiedNFW1997ApJ}) and a stellar disk (\citealt{MiyamotoNagai1975PASJ}) are shown in \textit{dashed brown} and \textit{dot-dashed green} lines respectively. The \textit{solid purple} line shows the total circular speed in the disk plane. See Section \ref{subsec:TempFieldInitialization} for the underlying potential and the right panel of Figure \ref{fig:equilibrium profile} for the azimuthal velocity of the ISM in this potential.} 
        \label{fig:gasVelocityProfile}
    \end{figure}

    With this choice of temperature and gravitational potential, $v_{\phi}$ (Eq. \ref{eq:RotVel2}) is non-zero at heights where the integral includes the transition in temperature from CGM to ISM along the vertical direction. 
    \adchange{This is because,} transitioning from $-\infty$ to the disk region, 
    the gradient in $c_s$ is steeper along $z$ than that along $R$. 
    \adchange{Hence the first term in parentheses contributes dominantly. 
    The azimuthal velocity quickly transitions to zero outside the disk because the temperature is uniform in the three components (ISM, CGM, ICM) and $\partial/\partial z^\prime$ and $\partial/\partial R$ in the integrand of Eq. \ref{eq:RotVel2} vanish. 
    Additionally, the potential and temperature are spherically symmetric, 
    leading to a zero velocity everywhere beyond the disk.} 
    
\begin{figure*}      
    \includegraphics[width=1.0\textwidth]{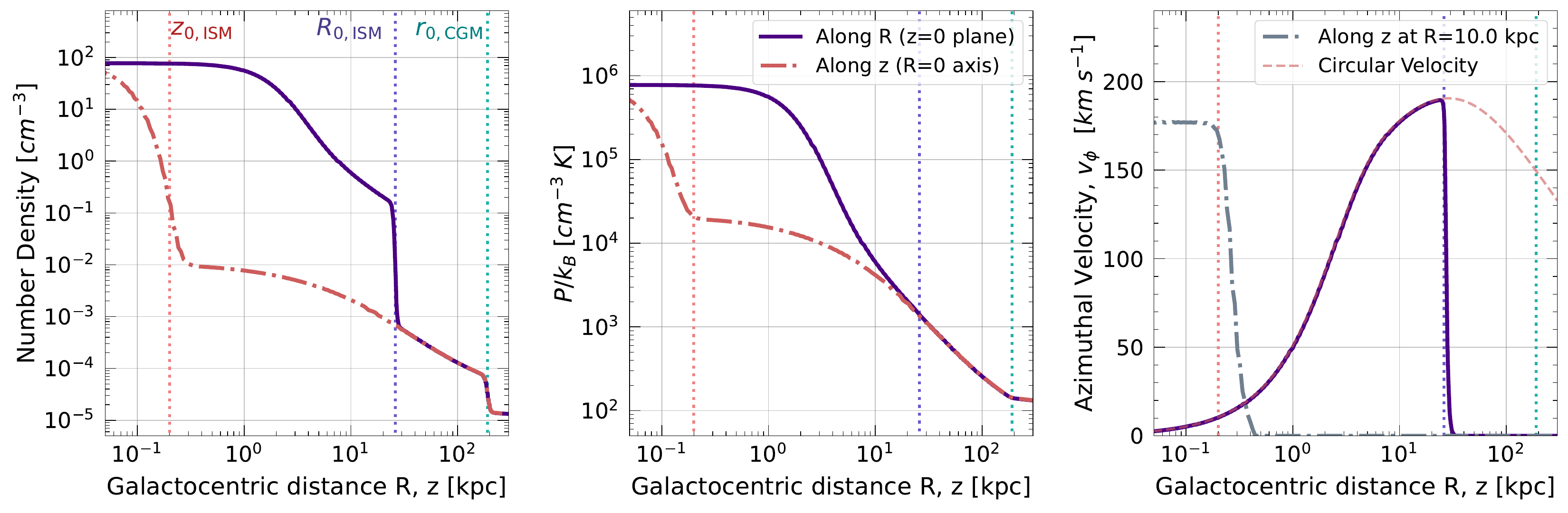}
    \caption{Equilibrium profiles for our model galaxy with respect to galactocentric distances $R$ and $z$. The vertical dotted lines mark the transition regions, similar to Figure \ref{fig:temperature profile}. \textit{Left panel:} number density; 
    \textit{Middle panel:} pressure; 
    \textit{Right panel:} azimuthal velocity $v_{\phi}$ of galactic gas distribution. In each panel, the \textit{solid violet} line shows the profile along $R$ in the disk plane (at $z=0$) and the \textit{dot-dashed red} \adchange{line} is along $z$ (at $R=0$ axis).
    The azimuthal velocity of the gas in the right panel (\textit{solid violet} line) is close to the circular speed \adchange{in} the disk plane (\textit{pink dashed} line) and is obtained numerically by solving Eq. \ref{eq:RotVel2} (see  point \ref{vphi-discussion} in Section \ref{sec:Framework} and Eq. \ref{eq:vphi_circular}). The fall in circular speed beyond $\sim 40\ \rm kpc$ (dashed line) is according to the underlying NFW dark matter halo ($\propto \ln(r)/r$) at large radii. The \textit{gray dot-dashed} line in the right panel shows the azimuthal velocity along z- at $R=10\ \rm kpc$. 
    }
    \label{fig:equilibrium profile}
\end{figure*}

    Figure \ref{fig:gasVelocityProfile} shows the circular speed in the disk plane due to the gravitational potential from the dark matter halo and the stellar disk. The maximum \adchange{circular speed} for our model galaxy (with mass and size comparable to JO201/JO206; see Table \ref{tab:Fiducial galaxy params}) is $\sim 190\ \rm km \ s^{-1}$ and is dominated by the NFW potential at large distances. Note that the azimuthal velocity of the gas in the disk plane (cf. right panel of Figure \ref{fig:equilibrium profile}; Eq. \ref{eq:RotVel2}) is comparable to the circular speed, i.e.,
    \begin{equation}\label{eq:vphi_circular}
        v_{\phi,\;\rm ISM} (R,0) \approx 
        v_c \equiv \sqrt{R\frac{\partial \Phi}{\partial R}\Bigr|_{z=0}}\;\;. 
    \end{equation}
    
    Figure \ref{fig:equilibrium profile} shows the profiles of number density, pressure, and azimuthal velocity of the gas along the radial and vertical directions from the center of the galaxy.
    \adchange{Since the prescribed gravitational acceleration of the satellite is weak beyond its
    virial radius, the ICM 
    density profile in the leftmost panel is 
    flat. The resulting ICM density is very small ($\sim 10^{-5}\ \rm cm^{-3}$), which is expected since the gravitational potential of the cluster is not included in our modeling. Without accounting for the cluster potential, the hot gas at $10^7 \ \rm K$ cannot be sustained at a higher density. The central panel demonstrates that our equilibrium solution provides a smooth transition in pressure from the ISM to the surrounding CGM,  and eventually to the ICM beyond the virial radius of the galaxy. In the rightmost panel, the solid purple line shows the azimuthal velocity of the gas in the disk plane, which declines to zero beyond the outer edge of the disk $R_{\rm 0, ISM}$\ ($26 \rm \ kpc$). The gray dashed line shows the azimuthal 
    velocity perpendicular to the disk plane (along $z$) at a distance $R=10\rm \ kpc$ from the center. The velocity along the vertical direction again vanishes beyond the ISM-CGM interface at $z_{\rm 0, ISM}$ ($0.2 \rm \ kpc$).
    }
    In contrast to the quasi-steady solutions from several previous studies, our profiles are steady-state solutions to the fluid equations. 
    When initialized in isolation, the galaxy remains stable for $5\rm \ Gyr$ in our simulations.
    
\begin{table*}
  \centering
  \caption{Fiducial parameters for our model galaxy and its surrounding environment. Known observational constraints are compiled in the two rightmost columns for \textit{jellyfish galaxies} JO201 and JO206 (\citealt{Ramatsoku2020A&A}).${}^{\dag}$}
    \label{tab:Fiducial galaxy params}
  \begin{tabular}{lcccc} 
    \hline
    Parameter & Description & Our Model & JO201 & J\rfchange{O}206\\
    \hline
        \smallskip
        \textbf{Stellar Disk} \\
        $M_{*} (M_{\odot})$ &  Stellar mass         & $3.98\times10^{10}$ & $3.6\times10^{10}$ & $8.5\times10^{10}$ \\
        $a_{*}$ (kpc)       & Radial scale height   & 4.0                 & -                  &-\\
        \smallskip
        $b_{*}$ (kpc)       & Vertical scale height & 0.2                 & -                  &-\\
        \smallskip
        \textbf{Gaseous Disk / HI Estimates} \\
        $M_{\rm HI} (M_{\odot})$           & Observed total HI mass               &  -                &  $1.65\times10^{9}$&  $3.2\times10^{9}$\\
        HI deficiency                  & 
        \begin{tabular}{@{}c@{}}Missing HI mass compared to \\ the THINGS field ${}^{\ddag}$ \end{tabular}
          &  -                & $60\%$             & $50\%$\\
        $M_{\rm gas,0} (M_{\odot})$  & 
        \begin{tabular}{@{}c@{}} Initial ISM mass (our model) or \\ the expected HI mass (obs.) ${}^{\S}$    \end{tabular}   & $6.7\times10^{9}$ &  $4.2\times10^{9}$ &  $6.4\times10^{9}$\\
        HI in tail $(M_{\odot})$       & \begin{tabular}{@{}c@{}} Tracer based ISM mass in tail at \\ t=2 Gyr (fid. sim.)/obs. tail mass      \end{tabular}  & $1.34\times 10^9$        &  $0.4\times10^{9}$ &  $1.8\times10^{9}$\\
        $T_{\rm ISM}$  (K)           & ISM temperature          & $10^4$     & -                  & - \\
        $z_{0, \rm ISM}$ (kpc)             & Vertical scale height          & 0.2               & -                  & - \\
        $R_{0, \rm ISM}$ (kpc)             & Radial scale height            & 25.0              & 24.6                 & - \\
        $\sigma_{z, \rm ISM}$ (kpc)        & Smoothening scale along z      & 0.05              & -                  & - \\
        \smallskip
        $\sigma_{R, \rm ISM}$ (kpc)        & Smoothening scale along R   & 1.0               & -                  & - \\
        \smallskip
        \textbf{Dark Matter Halo} \\
    $M_{\rm vir} ( M_{\odot})$         & Virial mass (or $M_{\rm 200}$)  & $10^{12}$         & -                  & - \\
        $c_{\rm 200}$                      & Concentration parameter     & 12                & -                  & - \\
        $H_0$ ($\rm km\ s^{-1}\ Mpc{}^{-1}$)& Hubble constant             & 67.4              & -                  & - \\
        $d_{\rm halo}$ (kpc)               & Core radius                 & 6.0               & -                  & -  \\
        \smallskip
        $r_{\rm 200}$ (kpc)                & Radius enclosing $M_{\rm 200}$  & 212.0             & -                  & - \\
        \smallskip
        \textbf{Circumgalactic Gas} \\
        $M_{\rm CGM} (M_{\odot})$    & CGM mass enclosing $r_{\rm 200}$ & $5.4\times10^{10}$  & -                & - \\
        $T_{\rm CGM}$ (K)       & CGM temperature                   & $2.0\times10^6$   & -                & - \\
        $r_{0, \rm CGM}$             & Transition from CGM to ICM        & $0.9\,r_{\rm 200}$      & -                & - \\
        \smallskip
        $\sigma_{r, \rm CGM}$ (kpc)  & Smoothening scale CGM to ICM      & 10.0                & -                & - \\
        \textbf{ICM wind properties} \\
        $n_{\rm ICM} (\rm cm^{-3})$     & Density at projected distance      & $[2\times 10^{-5},1\times 10^{-2}] ^{*}$      & $2\times10^{-3}$  & $2.65\times10^{-3}$\\
        $v_{\rm ICM} (\rm km\ s^{-1})$        & Velocity at projected distance     & $[400,3151] ^{*}$      & $3.34\times10^{3}$  & $1.25\times10^{3}$\\
        $P_{\rm ram} (\rm dyn\ cm^{-2})$   & Ram pressure at projected distance      & $[\rfchange{1.30\times10^{-13},3.03\times10^{-10}}] ^{*}$    & $2.2\times10^{-10}$ & $4.2\times10^{-11}$\\
        \hline
  \end{tabular}
  \begin{tablenotes}
      \item ${}^{\dag}$ {\fontsize{7pt}{7.5pt}\selectfont - indicates unavailable/ambiguous $\hspace{16em}{}^{\ddag}$ \citealt{Bigiel2008}}
      \item ${}^{\S}$ {\fontsize{7pt}{7.5pt}\selectfont Our modeled \rfchange{ISM} mass is comparable to the observed HI mass ($M_{\rm HI,0}$)$\; \;\;\;\;{}^{*}$ Range of values used in our suite of simulations.}
  \end{tablenotes}
\end{table*}

\section{Simulation Setup} \label{sec:simulationSetup}
Our simulations are carried out using \texttt{PLUTO} (v4.4p2), a conservative hydrodynamic code with static grid, developed by \citealt{PLUTO}. \adchange{The RPS simulations are performed} in 3-dimensional Polar geometry $(R, \phi, z)$.
Using a Polar grid over a Cartesian one has two-fold advantages: (i) it aligns with the natural geometry of the rotating disk, (ii) a static mesh can be selectively refined such that the ISM is well-resolved in the radial and vertical directions despite a coarse resolution in the azimuthal direction. \adchange{We use the HLLC (Harten, Lax, van Leer Contact) solver, with RK2 time-stepping and a linear reconstruction scheme, with a CFL number of 0.3. 
}
    
Our model galaxy is initialized at the center of the simulation domain (see Figure \ref{fig:equilibrium profile} and Section \ref{sec:Framework}; \adchange{for} fiducial parameters \adchange{see} Table \ref{tab:Fiducial galaxy params}). The minimum of the gravitational potential stays at $(R,z)=(0,0)\ \rm kpc$. The simulation box extends to 240 kpc in the radial direction and $240$ kpc on either side perpendicular to the disk plane. To simulate the ram pressure experienced by the galaxy, an additional velocity in the $\hat{z}$-direction is introduced to an otherwise stationary ICM.

We next discuss the \adchange{different} 
simulations that we perform to either model galaxy stripping or benchmark the underlying physical effect using control setups.
        
    \subsection{Stripping with gravity}\label{subsec:Grav}
    There is a wide range in density and temperature of the ICM, as well as a galaxy's infall velocity depending on its distance from the center of the host cluster. 
    Although an orbiting galaxy would experience varying ICM speed and density \adchange{in its trajectory, we} assign a uniform density and velocity to the ICM beyond the virial radius of the galaxy. 
    \adchange{This not only provides the simplest baseline scenario for stripping simulations but also enables us to explore the different stripping regimes unambiguously.}
    \rfchange{We, however, span a range of relative velocities - $400$ to $3200\ \rm km\ s^{-1}$. For the ICM density, we consider a range of values that represent densities in cluster outskirts ($\sim 10^{-5}\ \rm cm^{-3}$) to small cluster-centric distances($\sim 10^{-2}\ \rm cm^{-3}$).}  
    
    Our equilibrium solution from Section \ref{subsec:equilibriumSoln}, however, yields a small number density ($\sim 10^{-5}\ \rm cm^{-3}$) for the ICM held at $\sim 10^7$ K. Such small density can at best represent conditions in cluster outskirts or the IGM, but certainly not within $\sim 0.6\ r_{\rm 200}$ of the cluster (\adchange{ where number densities can be as high as $10^{-3}-10^{-1}\ \rm cm^{-3}$; \citealt{White1980ApJ, Canizares1983ApJ, perseusFabian1981ApJ, eRositaStacked2023MNRAS}}). 
    Therefore, to model a denser ICM representing stripping conditions at smaller cluster-centric distances, we have adopted two approaches to 
    \textit{elevate the ICM density} to a desired level:
    \begin{enumerate}[wide]
        \item \textit{\adchange{Increasing the ICM density} keeping the \adchange{ICM} temperature unaltered from the model prescription}. This, however, \adchange{increases the ICM pressure} compared to the equilibrium solution. 
        \adchange{This scenario is closer to reality given the observed range of temperatures and densities of the CGM and ICM. 
        We label these simulations as \textbf{OG} runs (\textit{O}verpressurized simulations with \textit{G}ravity).}
        
        \item \textit{\adchange{Increasing the ICM density} keeping the \rfchange{thermal pressure in the ICM} unaltered from the equilibrium solution}. This however, has the \adchange{problem} that the ICM temperature would accordingly decrease. 
        These simulations are labeled as \textbf{PG} runs (\textit{P}ressure equilibrium with \textit{G}ravity). \rfchange{Although the ICM temperature in these setups does not represent typical ICM conditions, we use these runs to evaluate the relative role of thermal over-pressurization (in OG runs) versus ram pressure on the evolution of a galaxy undergoing RPS (elaborated further in Section \ref{subsubsec:cgm_vs_ism}).}
    \end{enumerate}

    In this work, we only focus on the system's hydrodynamic evolution without radiative cooling and self-gravity of the gas. 
    \adchange{
    Our ram pressure $\rho_{\rm ICM} v^2_{\rm rel}$ varies between $\sim 10^{-13}-\ 3\times10^{-10}\ \rm dyn\ cm^{-2}$, the CGM-ICM density 
    contrast $\chi$ ranges between $0.01$ and $6.0$, and the Mach number of the ICM wind $\mathcal{M}=v_{\rm rel}/c_{\rm s, ICM}$ ranges from $0.8$ to $3.3$.
    
    For our fiducial parameters, we consider the typical ICM density of $10^{-3}\ \rm cm^{-3}$ and a velocity of $800\ \rm km\ s^{-1}$. 
    We now discuss the underlying simulation grid and the initialization of the ICM wind within the domain. }
    \vspace{-7pt}
    
        \subsubsection{\adchange{Grid and geometry}}
        \label{subsec:gridding}
\begin{figure}
        \centering
        	\includegraphics[width=\columnwidth]{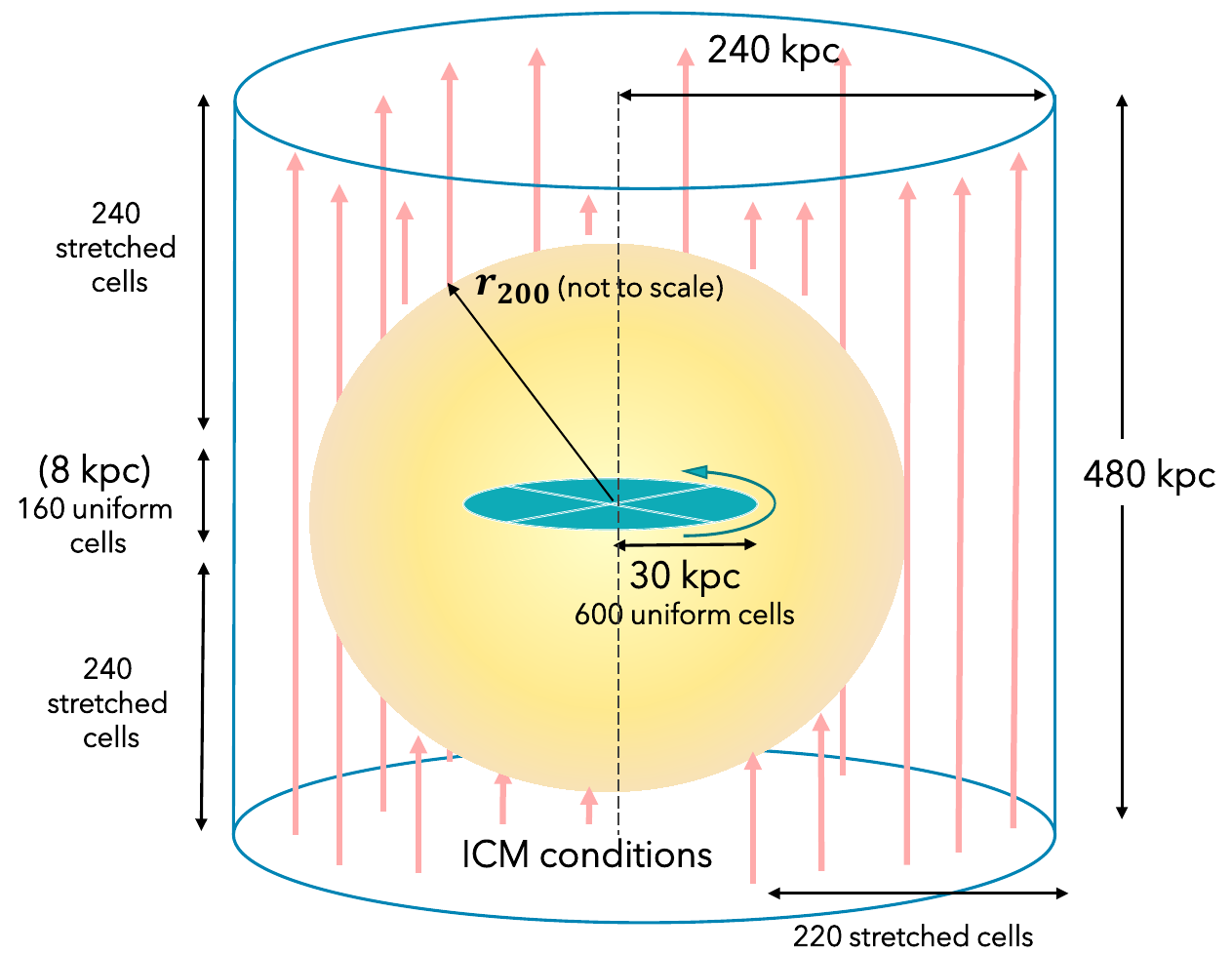}
            \caption{\adchange{Schematic representation of our simulation domain.} The galaxy is at the center of the domain with a fiducial resolution of ($N_R \times N_{\phi} \times N_z)\ \equiv (820 \times 50 \times 640$) in 3D polar coordinates $(R, \phi, z)$. The rotating disk is surrounded by a non-rotating circumgalactic medium. Beyond $r_{\rm 200}$, we initialize the ICM density, pressure\rfchange{,} and the relative velocity $v_{\rm rel}$ as illustrated \adchange{-in the region with arrows (the direction of wind)}. The rotating disk is resolved uniformly along $R$ by 600 grid cells within $30$ kpc and along $z$ by 160 grid cells within $4$ kpc on either side of the disk plane. Thus, the effective resolution for the disk is $50$ pc. Beyond this region, the grid is `stretched' \adchange{(grid size increases in geometric progression) with} 240 grid cells till 240 kpc along the vertical and 220 grid cells till $240$ kpc along the radial direction (see Section \ref{subsec:gridding} and Appendix \ref{app:grid}).
            }
            \label{fig:simulationBox}
        \end{figure}

        Our 3D-computational domain in Polar $(R, \phi, z)$ coordinates extends in the radial direction from $R_{\rm beg}=0.1$ kpc\footnote{\rfchange{Note that $R=0$ is outside the active simulation domain in Polar geometry as all cells, including the ghost cells, must have $R>0$.}} to $R_{\rm end}= 240$ kpc and in the vertical direction for 480 kpc. 
        Figure \ref{fig:simulationBox} is a cartoon illustrating our simulation setup and the grid arrangement. The galaxy is centered at $(R, z)= \rfchange{\rm (0,0)\ \rm kpc}$. 
        Along $R$, the grid is a combination of two patches - a uniformly spaced patch between $R_{\rm beg}=0.1\ \rm kpc$ and $R_{u}=30\ \rm kpc$ (for our fiducial run) with $N_{Ru}\adchange{=600}$ grid points. Beyond this, \adchange{we use a} stretched grids with $N_{Rs}\adchange{=220}$ number of grid points, such that the grid sizes increase in a geometric progression till $R_{\rm end}= 240$ kpc.
    
        Along the vertical \rfchange{axis}, the disk is resolved by $N_{zu}\adchange{=160}$ uniformly spaced points within $|z_{{\rm disk},u}|$ \rfchange{around} the center ($4\ \rm kpc$ for our fiducial run). 
        Beyond $|z_{{\rm disk},u}|$, we have $N_{zs}\adchange{=240}$ stretched grid points till the end of the domain $z_{\rm end}$.
        We use $N_{\phi}$ uniformly spaced grid points along the azimuthal direction. 
        So, the number of grid points in each direction are $N_R = N_{Ru}+N_{Rs}$, $N_{\phi}$  and $N_z =N_{zu} + 2N_{zs}$.
        Our fiducial run has a resolution of ($N_R \times N_{\phi} \times N_z)\ \equiv (820 \times  50 \times 640$) with the disk resolved by $50$pc (the smallest grid size in our simulation), while all other P(n)G/O(n)G runs have a resolution $(420\times 50 \times 280)$ with the disk resolved by $100\ \rm pc$ ($N_{Ru}\adchange{=300}$, $N_{Rs}\adchange{=120}$, $N_{zu}\adchange{=80}$, $N_{zs}\adchange{=100}$). Figure \ref{fig:stretched grid} shows how the cell width changes along the vertical and radial directions for the fiducial run (see also Appendix \ref{app:grid}).

        \subsubsection{Wind initialization in the domain}\label{windDomain}
        For all our simulations, we set the ICM wind properties, i.e. density $\rho_{\rm ICM}$, pressure $p_{\rm ICM}=\rho_{\rm ICM} k_B T_{\rm ICM}/(\mu m_p)$, and the relative velocity between the galaxy and ICM $v_{\rm rel}$, beyond $r_{\rm 200} (\sim 210\rm \ kpc$) for our model galaxy.
        This differs from previous works of ISM-ICM interaction, where the wind is usually \adchange{introduced through a boundary} 
        \citep{Roediger2007, Tonnesen&Bryan2010ApJ, Li2020MNRAS.492.1841L}. 
        Our wind initialization all over the domain is 
        justified because a satellite galaxy \adchange{moving through the host cluster, sees} the ICM gas moving towards it in the upwind direction and away from it down the wind.
        
        The transition in velocity starting from zero in the CGM to $v_{\rm rel}$ in the ICM is achieved by a gradually increasing velocity profile with 
            $v(r) =f(r)\times v_{\rm rel}$, where the smoothening factor
            \[
                f(r) = 0.5\times\left(1+\tanh \left(\frac{r -r_{0, \rm CGM}}{\sigma_{r, \rm CGM}}\right)\right)\; .
            \]
        Here, $r=\sqrt{R^2+z^2}$, and $\sigma_{\rfchange{ r, \rm CGM}}$ is the steepness with which the temperature changes across the CGM-ICM interface (see Table \ref{tab:Fiducial galaxy params}).
        
        \adchange{We maintain the wind conditions in the upwind direction (lower $\hat{z}$- boundary), while outflow conditions are imposed at the downwind (upper $\hat{z}$-) boundary. In the radial direction, the inner boundary ($R_{\rm beg}$) is assumed to be reflective, and the outer boundary ($R_{\rm end}$) is kept as outflow. Along the azimuthal direction $\phi$, we impose periodic boundary conditions.}

        \subsubsection{\adchange{Passive tracers for tracking CGM/ISM} \label{trackISMCGM}}
        To quantify the amount of stripped gas, we use passive tracers to define the mass of ISM/CGM within a chosen extent in the simulation domain as,
        \begin{equation}
           \label{eq:tracer_CGMISM}
            M_{\rm ISM/CGM}(t)= \bigintsss_{\mathit{V}}\rho C_{\rm ISM/CGM} dV, 
        \end{equation}
        where $C_{\rm ISM/CGM}$ represents the tracer value for ISM/CGM in any grid cell. The tracer initialization ($C_{\rm ISM}$/$C_{\rm CGM}$/$C_{\rm ICM}$) to mark the three components (ISM/CGM/ICM) is based on the initial temperature distribution. 
        
        As the galaxy experiences ram pressure from the ICM, as we will see in the following discussion, the ISM and the CGM evolve through two key processes. Primarily, the ICM transfers momentum to the CGM \adchange{via drag, which subsequently mediates the ram pressure on the ISM. Secondly,} the relative velocity \adchange{between the ICM, CGM and the ISM} causes shear-driven instabilities \adchange{and turbulent mixing across the three phases}.
        Although mixing of ICM and CGM produces \adchange{gas at} intermediate densities and temperatures, it does not destroy the bulk of the CGM (see Section \ref{subsubsec:cgm_vs_ism}). \adchange{Only when the mixed/perturbed CGM is dragged beyond the virial radius of the galaxy, it will no longer supply gas to the ISM and sustain star formation. Therefore, we quantify \textit{the CGM mass as the mass within $r_{\rm 200}$ of the galaxy} and \textit{similarly, the ISM mass as mass within its initial extent $(R,z)=(26, \pm 2)\ \rm kpc$}.}
        Stripping of ISM/CGM is thus quantified as \rfchange{the} reduction in $M_{\rm ISM/CGM}$ as \rfchange{original gas gets mixed and moves} out of these chosen extents. We also define a \textit{stripping timescale of the CGM}, $t_{\rm st, CGM}$, as the time by which only 10\% of the initial CGM mass remains within $r_{\rm 200}$ of the galaxy. This threshold value provides an unambiguous estimate of CGM stripping timescale from the trends in CGM evolution discussed later. 

    \subsection{Stripping without gravity}\label{subsec:noGrav}
    \adchange{To ascertain the relative importance of gravity in stripping the CGM versus the ISM, we devise and perform two sets of control simulations that resemble the `\textit{cloud-wind interaction}' problem.
    
    In the \textit{Control setup I} (Section \ref{subsec:control_setupCGM_disk}), we consider the evolution of the ISM+CGM of a galaxy facing an ICM wind in the absence of any gravitational potential of the galaxy. By comparing this idealized scenario against our RPS simulations, we can study similarities and differences in the wind-driven hydrodynamic interaction that disrupts and strips the diffuse CGM, unlike the dense ISM (protected by the gravitational potential of the galaxy). 
    
    To quantify the timescale of CGM stripping, we briefly digress to study a simplified problem -- the evolution of a uniform density cloud (representing the CGM; but with a uniform density $\sim$ average density of the CGM) embedded in a uniform wind (representing the ICM). We describe this set of simulations in \textit{Control setup II} (Section \ref{subsec:control_setupCGM_only}). Since a wide parameter space is relevant for the interaction of the CGM of satellite galaxies with the ambient ICM, our study complements the cloud crushing simulations that are} traditionally employed to study cold gas survival/entrainment in hot outflows (\citealt{Armillotta2017MNRAS.470..114A,GronkeOh2018MNRAS.480L.111G,Sparre2020MNRAS.499.4261S, Kanjilal2021MNRAS.501.1143K}). Typically, a galaxy would encounter varying background conditions of the ICM depending on its trajectory in the host cluster. Consequently, the CGM of a galaxy can be denser or lighter than the surrounding ICM. In the latter case, the CGM interaction with ICM wind would resemble the problem of \textit{interaction of a bubble in a dense wind}. In Control setup II, we will focus on this regime of cloud-wind interaction, and refer to it as the bubble drag regime. We discuss the salient features and differences between the two cloud-wind interaction regimes, namely, the classic cloud crushing regime (with density contrast $\chi > 1$) and the bubble drag regime ($\chi < 1$) later in Section \ref{subsec:TimeScaleBubble}. 
    \begin{figure}
        \centering\includegraphics[width=0.96\columnwidth]{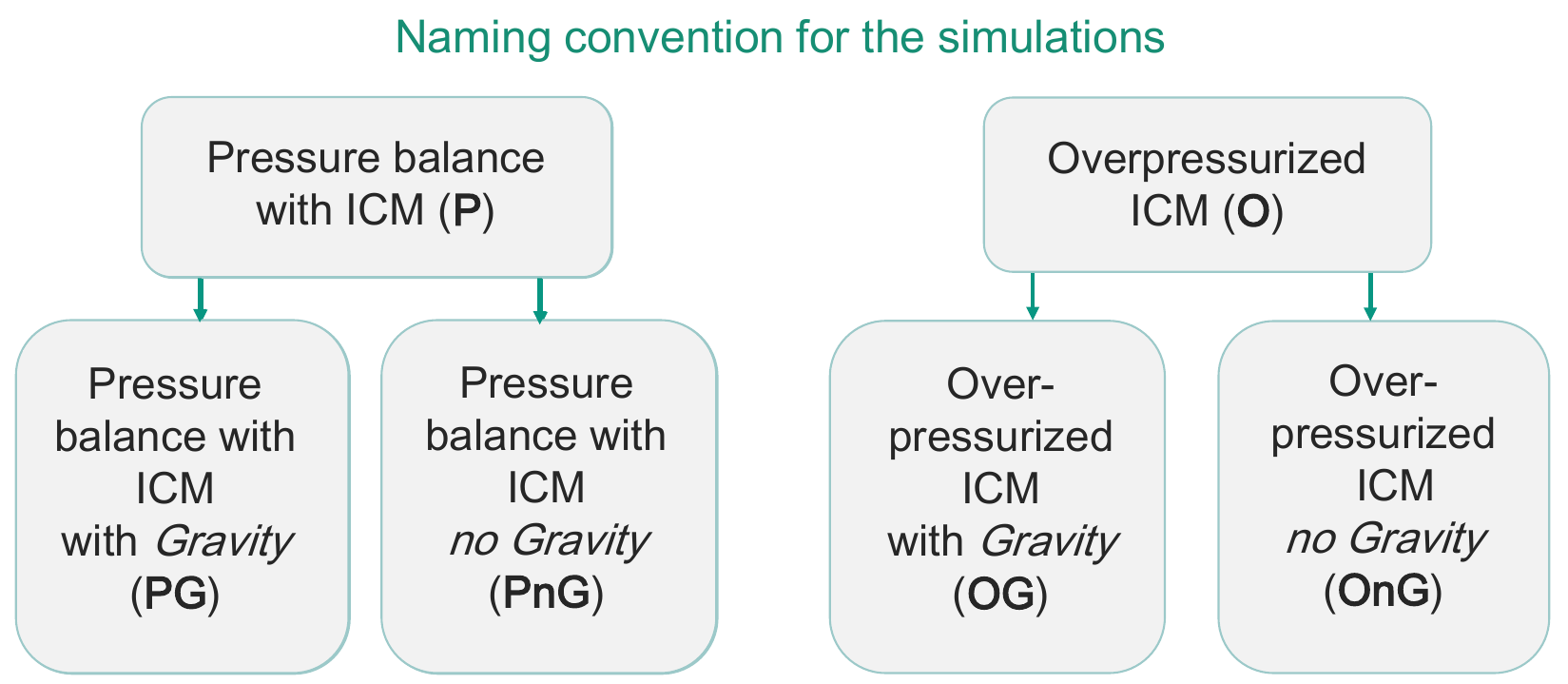}
        \caption{Naming convention for the suite of simulations in our work (see Sections \ref{subsec:Grav} \& \ref{subsec:noGrav}). The simulations are labeled according to these prefixes, followed by the ICM density and relative velocity values separated by the letter `v'. Hence the simulations are labeled as (PG/OG/PnG/OnG)($n_{\rm ICM}[\rm cm^{-3}]$)\text{v}($v_{\rm rel} [\rm km\ s^{-1}]$). }
        \label{fig:naming_conv}
    \end{figure}
        \subsubsection{\adchange{Control setup I: ISM+CGM in an ICM wind}\label{subsec:control_setupCGM_disk}} 
        \adchange{In these control simulations, we do not consider any gravitational potential (from the dark matter halo and the stellar disk of the galaxy) but take the density field directly from the RPS simulations described in Section \ref{subsec:Grav}. We initialize an equilibrium configuration with no rotational velocities in either the ISM or the CGM. The choice of such an initialization gives us the closest approximation to the gas distribution within the galaxy, \textit{when the gravity is `turned off'}. It is worth noting that in the absence of the galaxy's gravitational potential, Eqs. \ref{eq:equilibriumR} \& \ref{eq:EqbZ} suggest that the radial and vertical pressure gradients are respectively $\partial P/\partial R=\rho v_{\phi}^2/R$ and $\partial P/\partial z=0$. Therefore, the radial profile of $\rho v_{\phi}^2/R$ must be identical at all heights above and below the disk plane. Having a rotating ISM in this scenario would produce an unphysical rotation of the gas distribution out to large distances from the disk. We therefore have $v_{\phi}=0$ everywhere in our initialization.
        Thus, all the components (ISM+CGM and ICM) are non-rotating when the gravity is \textit{turned off}.
        Furthermore, the ISM and CGM are held at a constant pressure set to the pressure at the virial radius in the corresponding runs in our RPS simulations (PG/OG runs).}
        
        These control simulations are labeled as \textbf{OnG/PnG} depending on whether density fields are taken from OG/PG runs. Here, `\textbf{nG}' implies \textit{no-Gravity}. The isobaric condition is carried over to the ICM (beyond the virial radius of the galaxy) for the \textbf{PnG} runs. The \textbf{OnG} runs, on the other hand, have an identical thermal over-pressurization in the ICM as their OG counterparts. \adchange{The grid geometry, boundary conditions, and wind initialization are identical to the RPS runs, as described above in Section \ref{subsec:Grav}}.
        
        Figure \ref{fig:naming_conv} outlines the nomenclature followed in our suite of simulations. After the P(n)G / O(n)G tag, the simulation label has a number that corresponds to the number density (in \adchange{$\rm cm^{-3}$}) of the ICM followed by the letter `v' and the \adchange{value of relative velocity} (in $\rm km\ s^{-1}$) between the galaxy and the ICM. For instance, our fiducial run has the label `OG1e-3v800' which means it has an over-pressurized ICM with number density $10^{-3}\ \adchange{\rm cm^{-3}}$ and wind velocity $800\ \rm km\ s^{-1}$. The corresponding equilibrium run is labeled as `PG1e-3v800'. For these two simulations, the respective `no-Gravity' control setups would be labeled as `OnG1e-3v800' and `PnG1e-3v800'.

        \subsubsection{\adchange{Control setup II: The Bubble Drag problem}\label{subsec:control_setupCGM_only}}
        \adchange{
        We now describe our simpler adiabatic cloud-wind interaction simulations, designed to understand the CGM stripping timescale.
        We initialize a stationary spherical blob of uniform density (representing the CGM) with radius $R_{\rm cl}$ in pressure equilibrium with a uniform wind (representing the ICM) with velocity $v_{\rm wind}$ along the $\hat{z}$-direction. The blob is initialized at a density $n_{\rm cl}=10^{-4}\ \rm cm^{-3}$ ($\sim$ average CGM density).  
        Our simulations encompass a range of density contrast $\chi$ between the blob and the wind by varying the wind density  $n_{\rm wind}$ from $10^{-5}$ to $10^{-2}\rm \ cm^{-3}$. We refer to the blob as a \textit{cloud} or \textit{bubble} depending on whether it is \textit{denser} ($\chi$>1) or \textit{rarer} ($\chi<1$) than the ambient wind.\footnote{We use the standard notations from the cloud crushing literature; subscript 'cl' is used for a spherical blob (both a cloud and a bubble) and 'wind' for the wind. We appropriately replace 'cl' with 'b' whenever a distinction between the two regimes is necessary.}
        These simulations are carried out in 3D-Cartesian coordinates with a uniform grid. A box size of \rfchange{$(10, 10, 100)\ R_{\rm cl}$} for setups with $\chi > 1$ is required to capture the adiabatic evolution of the cloud along with its trail of mixed gas. On the other hand, for $\chi < 1$, a smaller box size of \rfchange{$(8, 8, 40)\ R_{\rm cl}$} is sufficient to contain the bubble and its wake. The initial blob, in both the cloud crushing and bubble drag problem, is uniformly resolved by 16 grid cells ($R_{\rm cl}/d_{\rm cell} = 16$; $d_{\rm cell}$ is the grid size). 
        }
        
        \adchange{We choose outflow boundary conditions in all the directions except for the lower $z$-boundary, which is fixed to the wind conditions. We mark the initial cloud/bubble with a tracer field $C$ (1 in the cloud/bubble and 0 in the wind).
        We track the velocity of the cloud/bubble in the wind direction as 
        \begin{equation}
            \langle v_{z, \rm \ cl/b} \rangle=\frac{\bigintsss \rho C_{\rm cl/b} v_z \ dV} { \bigintsss \rho C_{\rm cl/b} \ dV},
            \label{eq:vel_COM}
        \end{equation}
        where the subscript `cl' denotes cloud and `b' denotes bubble, and the integral includes the entire simulation domain.
        Using Eq. \ref{eq:vel_COM}, we adopt a cloud tracking scheme that continuously shifts to the center of mass of the blob following \citealt{Shin2008ApJ} and \citealt{Dutta_2019}. 
        The evolution in the adiabatic cloud-wind interaction problem only depends on the density contrast $\chi$ and the Mach number of the wind $\mathcal{M}_{\rm wind}=v_{\rm wind}/c_{\rm s, wind}$. 
        
        Table \ref{tab:bubble drag params} lists the range of densities and Mach number considered in our \textit{Control setup II}.
        All the simulations are carried out till $20\ t_{\rm cc}$, where $t_{\rm cc}$ is the cloud crushing time defined as the characteristic time for the growth of hydrodynamic instabilities. $t_{\rm cc}=\sqrt{\chi_{\rm eff}} R_{\rm cl}/v_{\rm wind}$, where $\chi_{\rm eff}$ is the density ratio of the denser and the rarer media ($\chi_{\rm eff}\geq 1)$. Therefore, $\chi_{\rm eff}=(\rho_{\rm cl}/\rho_{\rm wind})$ for a cloud and $(\rho_{\rm wind}/\rho_b)$ for a bubble. } 
        
\section{Results}\label{Results}
\adchange{In Section \ref{sec:strippingResults}, we show that CGM stripping is qualitatively different from that of the ISM, as gravity plays a negligible role in binding the former.
Further, we find that the CGM stripping shows a clear delineation into two distinct regimes (in Section \ref{subsec:CloudCrushingComparison}), depending on its density contrast with the surrounding ICM.
}
The underlying physical process is elucidated through an additional set of simulations of a cloud/bubble seeded in a wind tunnel. 
\begin{table}
    \centering
    \caption{Compilation of results from all the ISM+CGM stripping simulations. The survival timescale for the CGM of our galaxy $t_{\rm st, CGM}$ for different ram pressure strengths $P_{\rm ram}$ is calculated as the time by which only 10\% of the initial CGM mass remains within $r_{\rm 200}$ of the galaxy.}
    \begin{adjustbox}{width=1.02\columnwidth,right}
    \begin{tabular}{lcccc}
    \hline
    Simulation label & $P_{\rm ram} (\rm dyn \;cm^{-2})$ & $T_{\rm ICM} (K)$& $\chi^{\S}$ & $t_{\rm st, CGM}^{\ddag}$\\
         \hline
    \adchange{PG2e-5v800} & $1.30 \times 10^{-13}$ & $10^7$    & \adchange{6.0}  & \xmark \\
    \adchange{PG2e-5v1600}& $5.22 \times 10^{-13}$  & $10^7$   & \adchange{6.0}  & 1.45 Gyr \\
    PG1e-4v800    & $6.52 \times 10^{-13}$ & $1.25 \times 10^{6}$ & 1.24    & 1.22 Gyr \vspace{0.3em} \\
    PG1e-3v800    & $6.52 \times 10^{-12}$ & $1.25 \times 10^{5}$ & 0.12    & 671 Myr \vspace{0.3em} \\
    PG1e-2v800    & $6.52 \times 10^{-11}$ & $1.25 \times 10^{4}$ & 0.012    & 525 Myr  \vspace{0.2em}\\
    \hdashline[0.5pt/1pt]  \addlinespace[0.5ex]
    OG4e-4v400    & $6.52 \times 10^{-13}$ & $10^7$    & 0.3   & \xmark \\
    OG4e-6v4000   & $6.52 \times 10^{-13}$ & $10^7$    & 30.0   & 1.11 Gyr \\
    OG1e-4v800    & $6.52 \times 10^{-13}$ & $10^7$    & 1.2    & 1.21 Gyr \vspace{0.3em} \\ 
    OG1e-4v1600   & $2.61 \times 10^{-12}$ & $10^7$    & 1.2   & 625 Myr \\
    OG5e-4v800    & $3.26 \times 10^{-12}$ & $10^7$    & 0.24   & 684 Myr \\
    {\color{MidnightBlue} OG1e-3v800} ${}^{\dag}$ & {\color{MidnightBlue}$6.52 \times 10^{-12}$} & {\color{MidnightBlue}$10^7$}   & {\color{MidnightBlue}0.12} & {\color{MidnightBlue}610 Myr} \\
    OG4e-3v400    & $6.52 \times 10^{-12}$ & $10^7$    & 0.03   & 1.60 Gyr \\
    OG2.5e-4v1600 & $6.52 \times 10^{-12}$ & $10^7$    & 0.48   & 467 Myr  \vspace{0.3em} \\
    OG1e-3v1600   & $2.61 \times 10^{-11}$ & $10^7$    & 0.12   & 331 Myr \\
    OG5e-3v800    & $3.26 \times 10^{-11}$ & $10^7$    & 0.024   & 579 Myr \\
    OG1e-2v800    & $6.52 \times 10^{-11}$ & $10^7$    & 0.012   & 557 Myr \\
    OG6.4e-3v1000 & $6.52 \times 10^{-11}$ & $10^7$    & 0.019   & 411 Myr \\
    OG2.5e-3v1600 & $6.52 \times 10^{-11}$ & $10^7$    & 0.048   & 283 Myr
    \vspace{0.3em} \\
    OG5e-3v1600   & $1.30 \times 10^{-10}$ & $10^7$    & 0.024   & 257 Myr\\
    OG1e-2v1600   & $2.60 \times 10^{-10}$ & $10^7$    & 0.012   & 238 Myr \\
    OG3e-3v3151   & $3.03 \times 10^{-10}$ & $10^7$    & 0.04   & 151 Myr  \vspace{0.2em}\\
    \hdashline[0.5pt/1pt] \addlinespace[0.5ex] \hdashline[0.5pt/1pt] \addlinespace[1.0ex]
    PnG2e-5v800   & $1.30 \times 10^{-13}$ & $10^7$    & 6.00   & \xmark  \\
    PnG2e-5v1600  & $5.22 \times 10^{-13}$ & $10^7$    & 6.00   & 1.15 Gyr \vspace{0.3em} \\
    \hdashline[0.5pt/1pt] \addlinespace[0.5ex]
    OnG1e-4v800   & $6.52 \times 10^{-13}$ & $10^7$    & 1.2    & 1.06 Gyr \vspace{0.3em} \\
    OnG1e-4v1600  & $2.61 \times 10^{-12}$ & $10^7$    & 1.2    & 588 Myr \\
    OnG1e-3v800   & $6.52 \times 10^{-12}$ & $10^7$    & 0.12   & 596 Myr \vspace{0.3em} \\
    OnG1e-3v1600  & $2.61 \times 10^{-11}$ & $10^7$    & 0.12   & 319 Myr \\
    OnG5e-3v800   & $3.26 \times 10^{-11}$ & $10^7$    & 0.024  & 529 Myr \\
    OnG1e-2v800   & $6.52 \times 10^{-11}$ & $10^7$    & 0.012  & 512 Myr \vspace{0.3em}\\ 
    OnG5e-3v1600  & $1.30 \times 10^{-10}$ & $10^7$    & 0.024  & 242 Myr \\
    OnG1e-2v1600  & $2.60 \times 10^{-10}$ & $10^7$    & 0.012  & 225 Myr \\
    \hline
    \end{tabular}
    \end{adjustbox}
    \begin{tablenotes}
        \item [] ${}^{\dag}$ {\fontsize{7pt}{7.5pt} Fiducial parameters}
        \item [] ${}^{\S}$ {\fontsize{7pt}{7.5pt} Ratio of the average CGM density ($1.2 \times 10^{-4}\ \rm cm^{-3}$) to the ICM density}
        \item [] ${}^{\ddag}$ {\fontsize{7pt}{7.5pt} \xmark  $\;$ indicates that CGM survives until the end of simulation}
    \end{tablenotes}
    \label{tab:stripping_results}
\end{table} 

\begin{figure*}\label{Grid}
    \centering
    \includegraphics[width=0.97\textwidth]{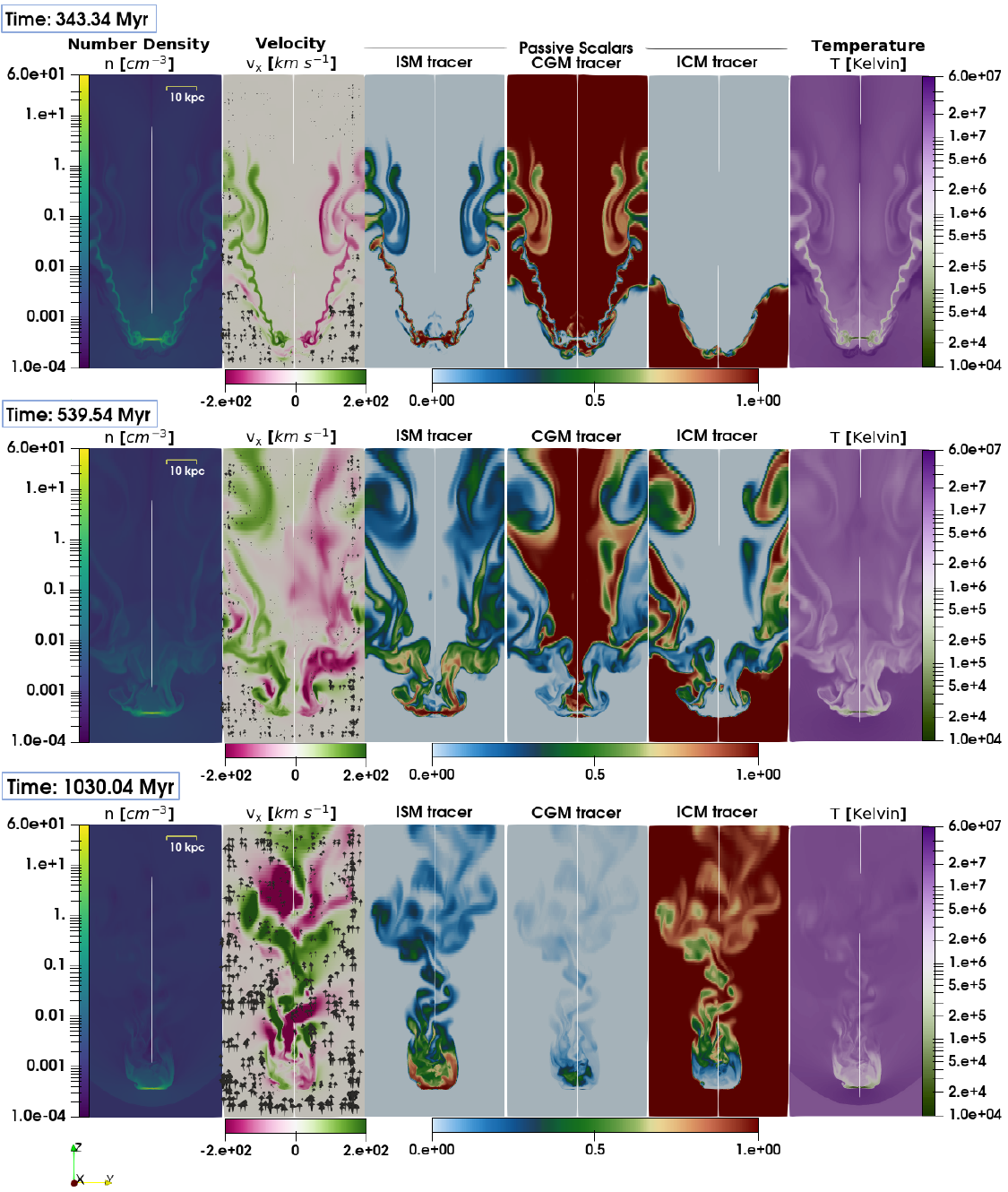}
    \caption{Slices through our simulation domain in the $yz$-plane at $x=0$ showing various fields for our fiducial simulation OG1e-3v800. Each panel shows a zoomed-in region (48 kpc$\times$100 kpc) around the disk, with the wind directed from bottom to top (along $\hat{z}$). \textit{Left to right}: The number density in logarithmic scale, velocity along the line of sight ($v_x$ for a slice in $yz-$plane), passive tracers for the \adchange{ISM}, CGM \& ICM (1 denotes \textit{pristine} tracer component), and temperature in logarithmic scale. \textit{Top to Bottom}: Evolution in time. The top row shows \adchange{the initial phase} when the CGM mediates the ram pressure to the ISM, stripping it `outside in'. The uplifted ISM mixes with the CGM, forming tails with rotation signatures carried over from the disk. In the second row, at $\sim 540$ Myr, the ISM directly interacts with the ICM. By this time, the bulk of the CGM has gained enough momentum and has started to get stripped (over a drag time $t_{\rm drag}=2\times r_{\rm CGM}/v_{\rm rel}\sim 500\ \rm Myr$; see Section \ref{subsec:CloudCrushingComparison} for a discussion on relevant timescales). The third row shows the continuous stripping of ISM by the ICM wind. A fraction of CGM remains close to the disk where the gravitational attraction is strong. \adchange{Analysis of mass evolution of the ISM and the CGM reveals that the ISM is ram pressure stripped, while the bulk of the CGM is stripped due to drag (see Section \ref{subsubsec:cgm_vs_ism} and Figure \ref{fig:cgm_vs_ism_mass}). The evolution is available at this video link: \href[pdfnewwindow=true]{https://youtu.be/9Sy54nsX5zs}{https://youtu.be/9Sy54nsX5zs}.}
    }
    \label{fig:rps_snapshots}
\end{figure*}

Table \ref{tab:stripping_results} lists the parameters from all the simulations in our suite, along with 
the measured CGM stripping timescale $t_{\rm st, CGM}$ (see Section \ref{trackISMCGM} for its definition).

    \subsection{Stripping of ISM \& CGM}\label{sec:strippingResults}
        \subsubsection{The fiducial run}\label{subsec:fiducial_run}
        Figure \ref{fig:rps_snapshots} shows various stages of RPS of the ISM and the CGM in our fiducial run (OG1e-3v800). All the panels are slices through \adchange{our 3D simulation domain in} the $yz$-plane, zoomed-in to show a ($\rfchange{48}$ kpc $\times$ $100$ kpc) region in the neighborhood of the disk. The ICM wind is directed along the $\hat{z}$-direction. 
        From left to right, \adchange{the panels show slice plots of} the number density (in log-scale), the velocity along the line of sight ($v_x$ for the $yz$-plane), the concentration of passive tracers for the ISM, CGM, and ICM, and the temperature (in log-scale).
        
        The top panel in Fig. \ref{fig:rps_snapshots} shows the snapshot at $\sim 340$ Myr.
        Although the ICM wind tracer \rfchange{(see \textit{fifth} column from the left) is yet to} reach the disk, the CGM is able to communicate the ram pressure to the ISM. \adchange{This is evident from the snapshots of density and ISM tracer, which shows large swaths of ISM already displaced from its initial position. Stripping of the ISM by ram pressure is at play here because it commences outside-in, as revealed by these snapshots.} The edge of the CGM-ICM interface has reached the disk by a crossing time ($r_{\rm CGM}/v_{\rm rel}\sim 300 \rm Myr$) due to efficient momentum transfer from the wind to the CGM. \adchange{We find that the CGM ahead of the disk is already co-moving with the wind as indicated by the arrows showing the instantaneous gas velocity (in \textit{second} panel).} Two distinct mixing layers can be visually identified - the ISM-CGM mixing layer and the CGM-ICM mixing layer. The CGM shadowed by the \rfchange{ISM} is still stationary, causing the stripped tail to fan out as it moves downstream. 
        \adchange{The stripped ISM carries the signature of disk rotation} for about $\sim 30$ kpc in the trailing wake.
        
        The second row shows that at $\sim 540\ \rm Myr$, the ISM starts interacting directly with the ICM, and some of the CGM is still shielded by the dense ISM. Down the wind, some of the \adchange{uplifted ISM} experiences shear, mixes, and attains intermediate temperatures. 
        
        In the third row, we see a `\textit{continuous stripping phase}', where the ISM continuously mixes with the ICM. As seen in the fourth column, a fraction of CGM gas still lingers close to the disk. This is because the gravity is strong enough to keep the CGM bound in the vicinity of the disk. The rotational signatures, as seen in the line of sight velocity 
        (see \textit{second} column from left; the colorbar is capped at $\pm200\ \rm km\ s^{-1}$),
        are correlated with regions with high ISM tracer for over $20\ \rm kpc$ down the tail. Beyond that, the rotational sense is lost as mixing introduces significant random velocities in the tail. 
        The line of sight velocity has alternating signatures of gas moving in and out of the $ yz$ plane, which are indicative of helical motions in the wake.
        Because of the absence of radiative cooling in this study, we find that the disk is slightly fluffier than its initial configuration, as shock-heating by the impinging wind causes a slight rise in disk temperature at the outskirts (see \textit{rightmost} column in lower panel). 
        
        The presence of CGM has two important consequences -- (i) the environment of the stripped ISM tails can have a significant contribution from the CGM and (ii) the uplifted ISM mixes with the CGM \adchange{and ICM}, forming tails with rotation signatures carried over from the disk over a few 10s of kpc. 

        \subsubsection{CGM versus ISM stripping}\label{subsubsec:cgm_vs_ism} 
\begin{figure}
            \centering
            \includegraphics[width=\columnwidth]{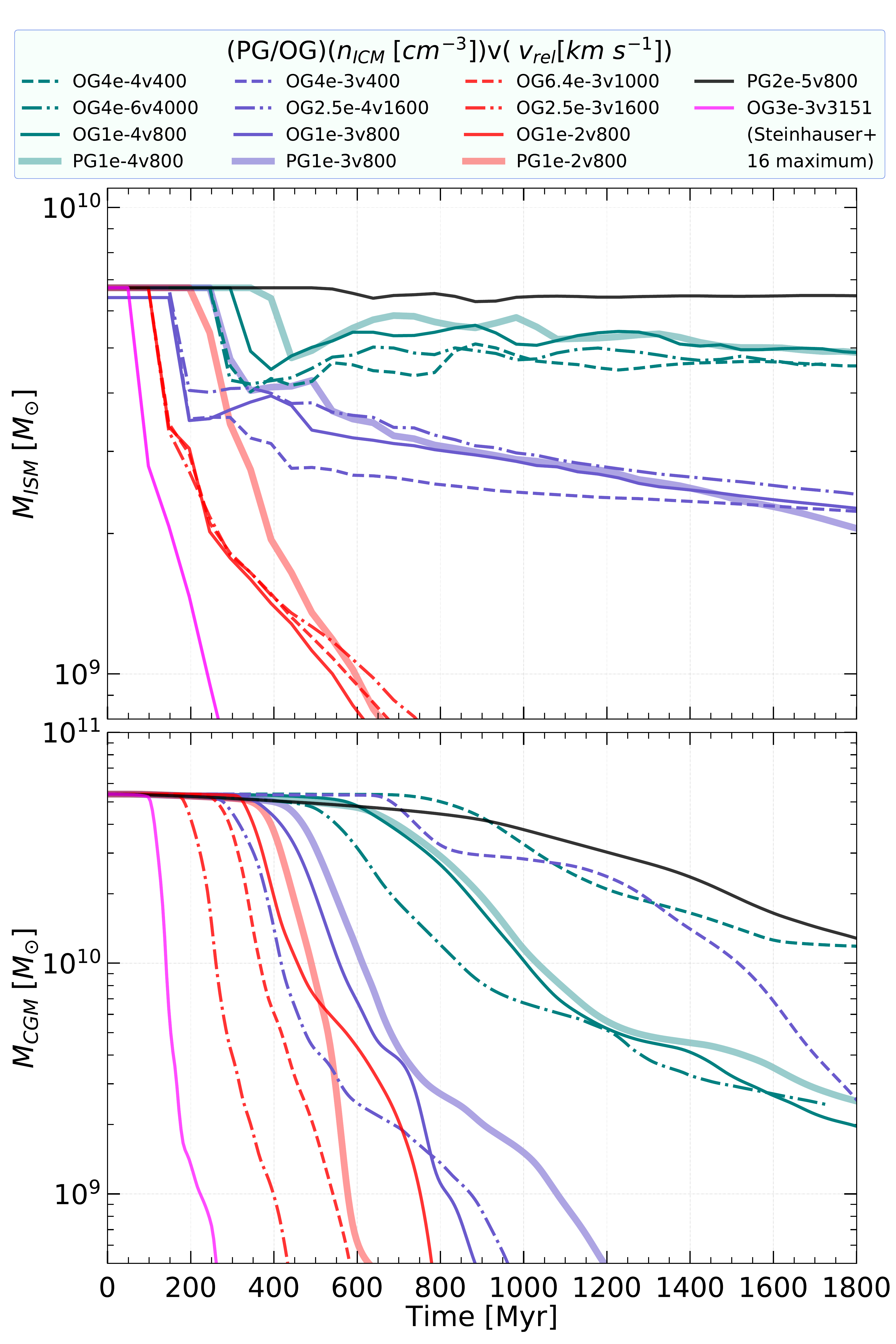}
            \caption{Evolution of the mass of the ISM $M_{\rm ISM}$ (\textit{top}) and the CGM $M_{\rm CGM}$ (\textit{bottom}). 
            The lines are colored according to the strength of the impinging ram pressure $P_{\rm ram} = \rho_{\rm ICM} v_{\rm rel}^2$ $=\ \left[\rfchange{1.30}\times10^{-13},\right.$
            $ \rfchange{6.52}\times10^{-13},\ $ 
            $ \rfchange{6.52}\times10^{-12},\ $ 
            $ \rfchange{6.52}\times10^{-11},\ $ 
            $ \left.\rfchange{3.03}\times10^{-10} \right]$
            $ \rm dyn\ cm^{-2}$ for [\rfchange{\textit{black, green, violet, red, magenta}}] lines (see Figure \ref{fig:naming_conv} for the naming convention used in the legend).
            The ISM evolution is found to be entirely dependent on the ram pressure $P_{\rm ram}$ (consistent with the \citealt{GunnGott1972ApJ} criterion; see Section \ref{subsec:compare_prev_work}), as lines 
            in the same color show similar evolution. 
            \rfchange{The CGM mass evolution, however, is not similar for the same ram pressure} \adchange{as lines with the same color are widely separated in the bottom panel}. The mass evolution for the fiducial run OG1e-3v800 is shown by the \textit{thin solid violet} line (see Figure \ref{fig:rps_snapshots} \& Section \ref{subsec:fiducial_run}). 
            The stripping progresses the fastest for the highest ram pressure shown by the \textit{magenta} line. In contrast, for some cases, a significant mass of the CGM and the ISM survive for more than even 2 Gyr (see Table \ref{tab:stripping_results}).
            }
            \label{fig:cgm_vs_ism_mass}
        \end{figure}

        The top panel of Figure \ref{fig:cgm_vs_ism_mass} shows the evolution of the ISM mass $M_{\rm ISM}(t)$ \adchange{(see Eq. \ref{eq:tracer_CGMISM} for its definition)}. 
        Lines with \adchange{the same color show runs with the same ram pressure. For the same line color (ram pressure), the different linestyles denote different combinations of ICM density and relative velocity.}
        
        \rfchange{According to the classic Gunn \& Gott criterion (\citealt{GunnGott1972ApJ}), gas from the ISM is stripped when the impinging ram pressure ($\rho_{\rm ICM}v_{\rm rel}^2$) exceeds the gravitational restoring pressure from the stellar and dark matter potential of the galaxy (which is $2\pi G \Sigma_{*} \Sigma_{\rm gas}$ for an infinitely thin disk; $\Sigma_{*, \rm gas}$ is the surface density of stars+dark matter, gas).
        With increasing ram pressure strength from $P_{\rm ram}= 1.30\times 10^{-13}\ \rm dyn \ cm^{-2}$ in black (\adchange{PG2e-5v800}) to $P_{\rm ram}= 6.49\times 10^{-10}\ \rm dyn \ cm^{-2}$ in magenta, the ISM mass progressively becomes lower. 
        The ISM mass evolution is determined solely by the ram pressure as a similar evolution of $M_{\rm ISM}(t)$ is seen for the lines in the same color. The ISM is thus ram \adchange{pressure} stripped as per the Gunn \& Gott criterion (see Figure \ref{fig:mcCarthy estimates} and Section \ref{subsec:compare_prev_work} for quantitative details on the `modified' Gunn \& Gott criterion for 
        our ISM+CGM simulations).} 
        However, slight differences in the mass evolution among the same ram pressure runs do arise from shock heating of the ISM gas at higher densities of the ICM wind, which alters $\Sigma_{\rfchange{\rm gas}}$ in the Gunn \& Gott criterion.
        \begin{figure*}
            \centering
            \includegraphics[width=0.97\textwidth]{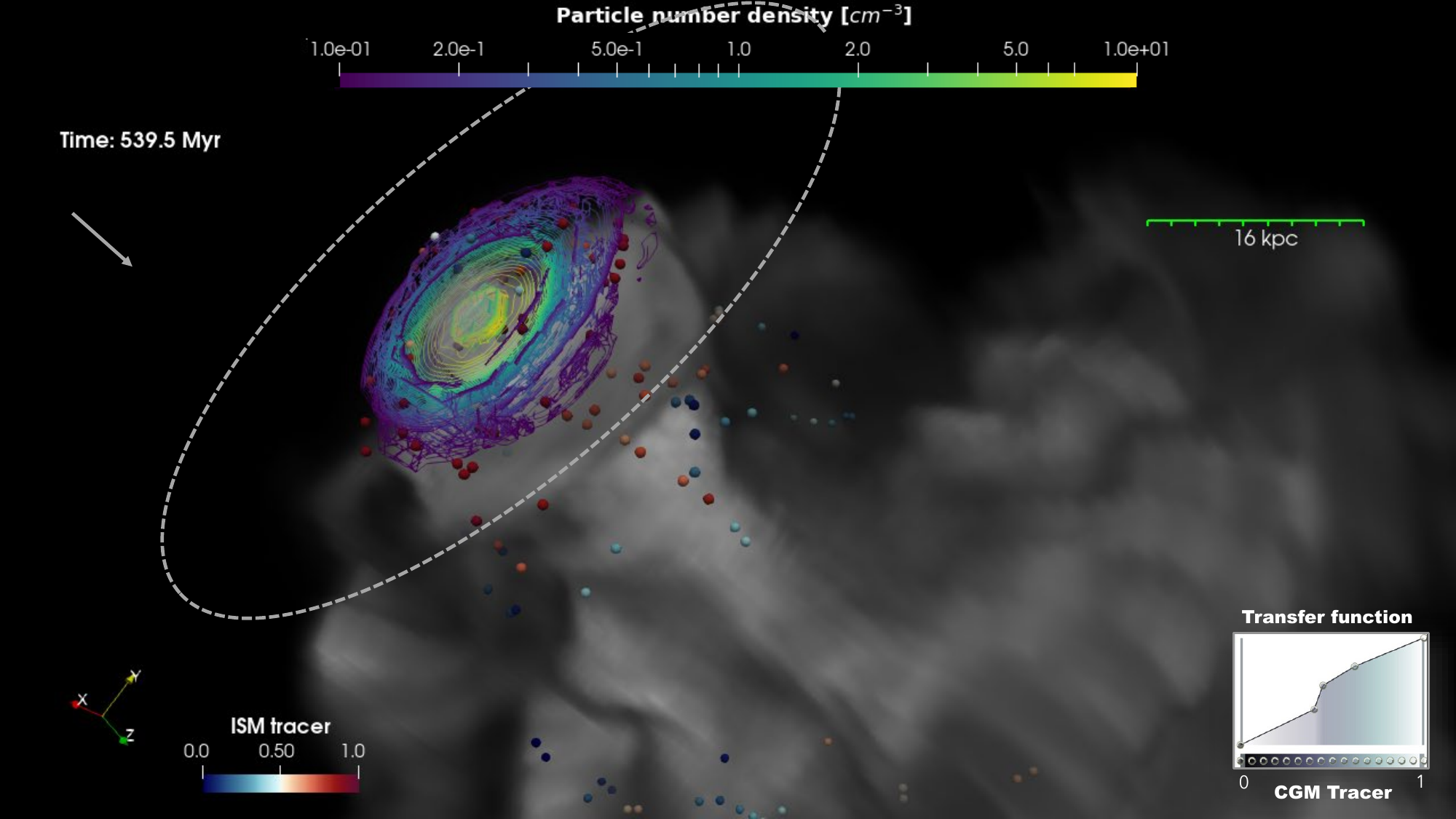}
            \caption{Volume rendering of different gas components of the galaxy in a zoomed\adchange{-in} region from our fiducial simulation OG1e-3v800, at 540 Myr. The dashed ellipse marks the initial extent of the disk. The ICM wind is along the $\hat{z}$-direction, as indicated by the white arrow (top left corner). At this time, both the ISM \& the CGM are getting stripped (cf. \textit{second row} of Figure \ref{fig:rps_snapshots}).
            The CGM stripping is revealed by the grayscale rendering of the CGM tracer using a transfer function shown in the inset (bottom right corner). 
            The disk is rendered as a wireframe and colored by its number density on a logarithmic scale (values indicated in the colorbar at the top). Lagrangian particles are introduced in the disk (in post-processing on \texttt{Paraview}) at time $t=0$, and their trajectories are calculated according to the evolving velocity field in the subsequent snapshots. The spheres mark the instantaneous position of these particles and are colored according to the ISM tracer at their location (see colorbar at the bottom left corner). Down the wind, the gradient in the color of these spheres indicates dilution of the ISM as it mixes with the CGM and the ICM.
            The complete evolution is available in this link: \href[pdfnewwindow=true]{https://youtu.be/CieB_IgCy4o}{https://youtu.be/CieB\_IgCy4o}.
            }
            \label{fig:volumeRendering}
        \end{figure*}

        In contrast, the evolution of the CGM mass \adchange{$M_{\rm CGM}(t)$, shown in the bottom panel of Figure \ref{fig:cgm_vs_ism_mass}, at the same ram pressure (same line color) is widely separated. This highlights that} instead of the ram pressure, CGM stripping is caused by a different physical effect. 
        CGM stripping is caused by momentum transfer from the ICM wind to the CGM \adchange{as we expect 
        the gravity 
        to be negligible for the bulk of the CGM. To test this hypothesis, we simulate equivalent simulations with the same density distribution as our model galaxies, but in the absence of gravity (see Section \ref{subsec:control_setupCGM_disk}). We postpone our discussion on the results and implications of these `control' simulations later to Section \ref{subsec:CloudCrushingComparison}.}
        
        The mass evolution for the fiducial run, i.e., OG1e-3v800, is shown in the \textit{thin solid violet lines} in both panels of Figure \ref{fig:cgm_vs_ism_mass}. The ISM stripping starts at about $\sim 190$ Myr, while the CGM is not appreciably displaced beyond the virial radius till $400$ Myr (see
        Figure \ref{fig:rps_snapshots} \& Section \ref{subsec:fiducial_run}). 
        \rfchange{The \textit{black} line shows the mass evolution for the smallest ram pressure strength PG2e-5v800 run ($P_{\rm ram}=3.24\times10^{-13}\ \rm dyn\ cm^{-2}$; $n_{\rm ICM}\sim 10^{-5}\ \rm cm^{-3}$ 
        and $v_{\rm rel}=800\ \rm km \ s^{-1}$). 
        This represents stripping conditions at cluster outskirts where we find that the CGM survival time is long compared to the fiducial scenario (cf. Table \ref{tab:stripping_results}). The highest ram pressure run OG3e-3v3151 shown in \textit{magenta}, on the other hand, provides a baseline comparison with the existing literature on high ram pressure stripping, e.g., \citealt{Steinhauser2016} (their galaxy G1a faces a maximum ram pressure of $\sim 3\times 10^{-10}\ \rm dyn \ cm^{-2}$), where the authors find that CGM stripping over $\sim 150-200 \ \rm Myr$ can quench star formation in the galaxy 
        by inhibiting its gas supply.} 
        
        \adchange{We explore the relative importance of thermal over-pressurization and ram pressure in the mass evolution of the ISM and the CGM.} All the \textit{thin} lines (in \rfchange{various linestyles}) in Figure \ref{fig:cgm_vs_ism_mass} show the mass evolution for the runs with a \rfchange{thermally over-pressurized ICM beyond the virial radius of the galaxy} (OG runs; see  Section \ref{fig:naming_conv} for naming convention). The thick lines correspond to the runs with pressure equilibrium (PG runs; counterparts of the same color in thin solid lines). We note that this thermal over-pressurization does not drastically change the evolution of both the ISM and the CGM (the mass evolution in the OG and PG runs is similar). However, the ISM stripping commences slightly earlier in the OG runs. Upon visual inspection of the snapshots, we find that the thermal over-pressurization in the OG runs leads to an initial compression of the CGM, which reduces the CGM extent. 
        Nevertheless, the thermal over-pressurization does not affect the rate of momentum transfer from the ICM to the CGM when compared to the PG runs.
        
        \subsubsection{Insights from 3D visualizations}\label{sec:3dViz}
        In order to understand the coupling of CGM-ISM evolution during stripping, we post-process our fiducial simulation OG1e-3v800 with \texttt{Paraview} (v5.11.0). Figure \ref{fig:volumeRendering} is a volume-rendering of the CGM tracer and the number density of gas within our simulated box at $\sim540\ \rm Myr$. 
        The CGM tracer is rendered in grayscale, while the \adchange{number density field is rendered as a wireframe to mark the remaining ISM (colored in log scale as shown by the top colorbar)}. The 
        extent of ISM stripping can be estimated from the decreased size of the disk compared to its initial extent indicated by the dashed ellipse. The arrow shows the direction of the ICM wind, which \adchange{has displaced the CGM and} pushed the ISM out of the disk plane. 
        \begin{figure*}
            \centering 
            \includegraphics[width=\textwidth]{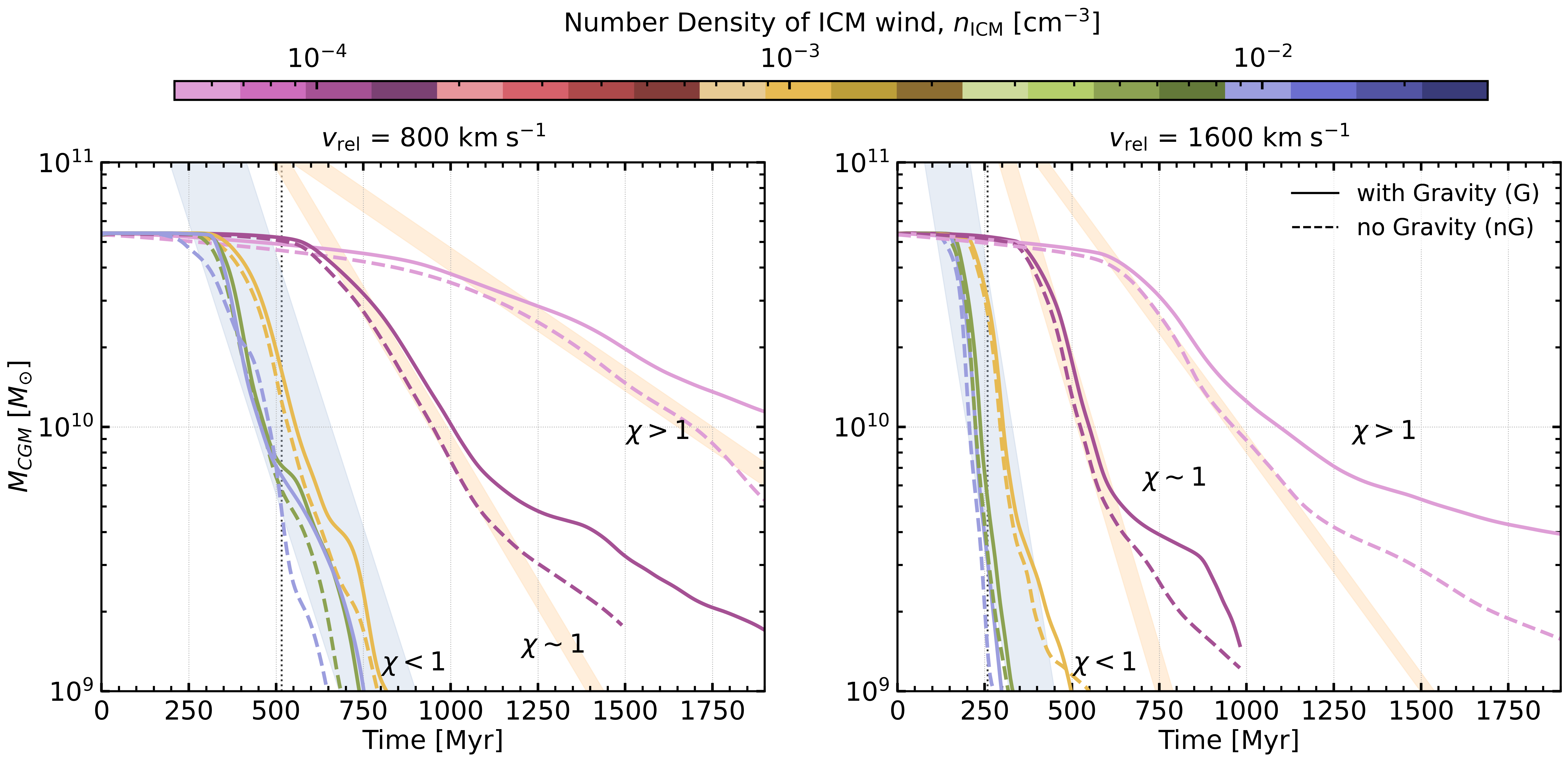}
            \caption{Time evolution of the CGM mass for various combinations of ICM number density $n_{\rm ICM}$ (indicated in the colorbar on a logarithmic scale) and relative velocity $v_{\rm rel}$;
            \textit{[Left, right] panel}: $v_{\rm rel}=800, 1600\  \rm km\ s^{-1}$. 
            The \textit{solid} lines are results from our ram pressure stripping simulations (PG/OG runs), and the \textit{dashed} lines in the same color are results from corresponding control simulation where \textit{`gravity is turned off'} (PnG/OnG runs). The similarity between the solid and dashed lines for the same density (color) implies that the CGM stripping in RPS simulations is akin to cloud crushing as gravity is negligible for the bulk of the CGM. The evolution separates in slopes, \rfchange{which} are grouped according to the density contrast of the CGM relative to the ICM wind (indicated by the shaded bands). The only PG runs shown here are those from our equilibrium modeling of the galaxy (the rightmost pink curves in both panels) having an ICM density $n_{\rm ICM}=2\times10^{-5}\ \rm cm^{-3}$. Runs with CGM to ICM density contrast $\chi>1$ show a different slope compared to all other values of $n_{\rm ICM}$ for which $\chi<1$. In between these two regimes, the magenta lines have an intermediate value of $\chi$, for which the slope is different. The slope of each band corresponds to a CGM mass loss rate that can be explained by a drag timescale $t_{\rm drag}$ (Eq. \ref{eq:mass_loss_rate} using results from idealized simulations of cloud-wind interaction, discussed further in Section \ref{subsec:TimeScaleBubble}).
            }
            \label{fig:nG_comp}
        \end{figure*}
        
        To understand the behavior of ISM uplifted from the disk, we introduced \rfchange{$6 \times 10^4$} particles in the zeroth snapshot of our fiducial simulation spread across the disk plane. These particles are tracked in the post-processing pipeline in \texttt{Paraview} as Lagrangian particles\footnote{Since Lagrangian particles are introduced in post-processing, quantitatively reliable trajectories for the \rfchange{particles} can only be \adchange{integrated from their initial position if the data dumping frequency is high}.} according to the local velocity field.
        These Lagrangian particles are shown as spheres in Figure \ref{fig:volumeRendering} and colored according to the ISM tracer value at their instantaneous position. There is a gradient in the color of the spheres down the wind, implying significant mixing of the ISM with both the CGM and the ICM. 
        \adchange{The gradient in ISM tracer and the presence of CGM tracer can manifest in the metallicity gradients observed in the tails of \adchange{jellyfish} galaxies (\citealt{Franchetto2020}).
        However, star formation in the tails and metallicity gradients within the ISM need to be carefully considered for a quantitative comparison, which is beyond the scope of our current work.}

    \subsection{CGM stripping versus cloud crushing}\label{subsec:CloudCrushingComparison}
    Being loosely bound to the galaxy, the gravitational effect (as compared to ram pressure) on the bulk of the CGM is negligible. Therefore, to gain insights into the CGM stripping timescales, we devise control setups akin to cloud crushing simulations (see Section \ref{subsec:noGrav}). 
    Since there are two types of simulations of ram pressure stripping (PG/OG) in our simulation suite, a fair comparative study of CGM stripping in `\textit{no-Gravity}' setups should also include two sets of control simulations, namely, \textbf{PnG} and \textbf{OnG}. 
        
        \subsubsection{\adchange{ISM+CGM in an ICM wind (Control Setup I)}}\label{subsec:ControlI: With Disk}
        Figure \ref{fig:nG_comp} \rfchange{shows} the CGM mass evolution for the RPS and no-Gravity simulations for \adchange{different values of relative velocity between the galaxy and the ICM (800 km s$^{-1}$ on the left, 1600 km s$^{-1}$ on the right panel)}.\footnote{A video of the evolution showing a comparison of the fiducial simulation with and without gravity can be found here: \clicks{https://youtu.be/NY1tiRyrO5M}{https://youtu.be/NY1tiRyrO5M}.} 
        The CGM mass in \textit{no-Gravity} setups (dashed lines) closely \rfchange{follows} the evolution of CGM mass in the equivalent RPS simulations (solid lines) for a long period. 
        This resemblance with cloud crushing simulations motivates our control \adchange{
        Setup II (see section \ref{subsec:TimeScaleBubble})}.
        We group the CGM mass evolution into categories based on the parameters of adiabatic cloud crushing -- (i) the relative velocity between the cloud (CGM here) and the wind (ICM here), and (ii) the density contrast $\chi$ between them.
        
        The evolution of the CGM mass (Eq. \ref{eq:tracer_CGMISM}) within the virial radius $r_{\rm 200}$ can be expressed in terms of a simple loss equation,
        \begin{equation}
            \label{eq:mass_loss_rate}
            \dot{M}_{\rm CGM}(t) = -\frac{M_{{\rm CGM}}(t)}{t_{\rm drag}},
        \end{equation}
        where $t_{\rm drag}$ is the drag time over which the CGM becomes co-moving with the background wind. \adchange{The shaded bands in Figure \ref{fig:nG_comp} show the solution to the above equation, with an offset in time set to match the fall of CGM mass (the thickness of the band just highlights the range of our simulations). The timescale $t_{\rm drag}$ is evaluated using the drag time from idealized simulations of cloud crushing or bubble drag (Control Setup II), which we discuss later in Section \ref{subsec:TimeScaleBubble}.}
        The dependence of CGM mass loss rate on the density contrast is evident from the grouping indicated by the three shaded bands with three different slopes in both the panels of Figure \ref{fig:nG_comp}. \adchange{In the left panel, the \textit{rightmost solid pink} curve (corresponding to lowest RPS strength PG2e-5v800) and the \textit{solid magenta} curve (for OG1e-4v800) have a density contrast $\chi \geq 1$. The yellow bands obtained with a drag time $t_{\rm drag}\sim \chi R_{\rm CGM}/v_{\rm rel}$, highlights that the stripping time scale is dependent on the density contrast for these cases. In contrast, all other values of $n_{\rm ICM}$ for which  $\chi < 1$ have the same slope independent of their density contrast, highlighted by the blue band. A similar grouping of $\chi<1$ runs is evident in the right panel where the wind velocity is higher ($v_{\rm rel}=1600 \ \rm km \ s^{-1}$). Therefore, in the $\chi<1$ regime, the diffuse CGM behaves like a bubble in the dense wind, with the drag time independent of its density contrast.}
        Moreover, for a higher relative velocity $v_{\rm rel}=1600 \ \rm km s^{-1}$ in the right panel, the slopes of CGM mass evolution are steeper, indicating a faster CGM stripping at high velocities. 
        \begin{figure*}
            \centering       
            \includegraphics[width=\textwidth]{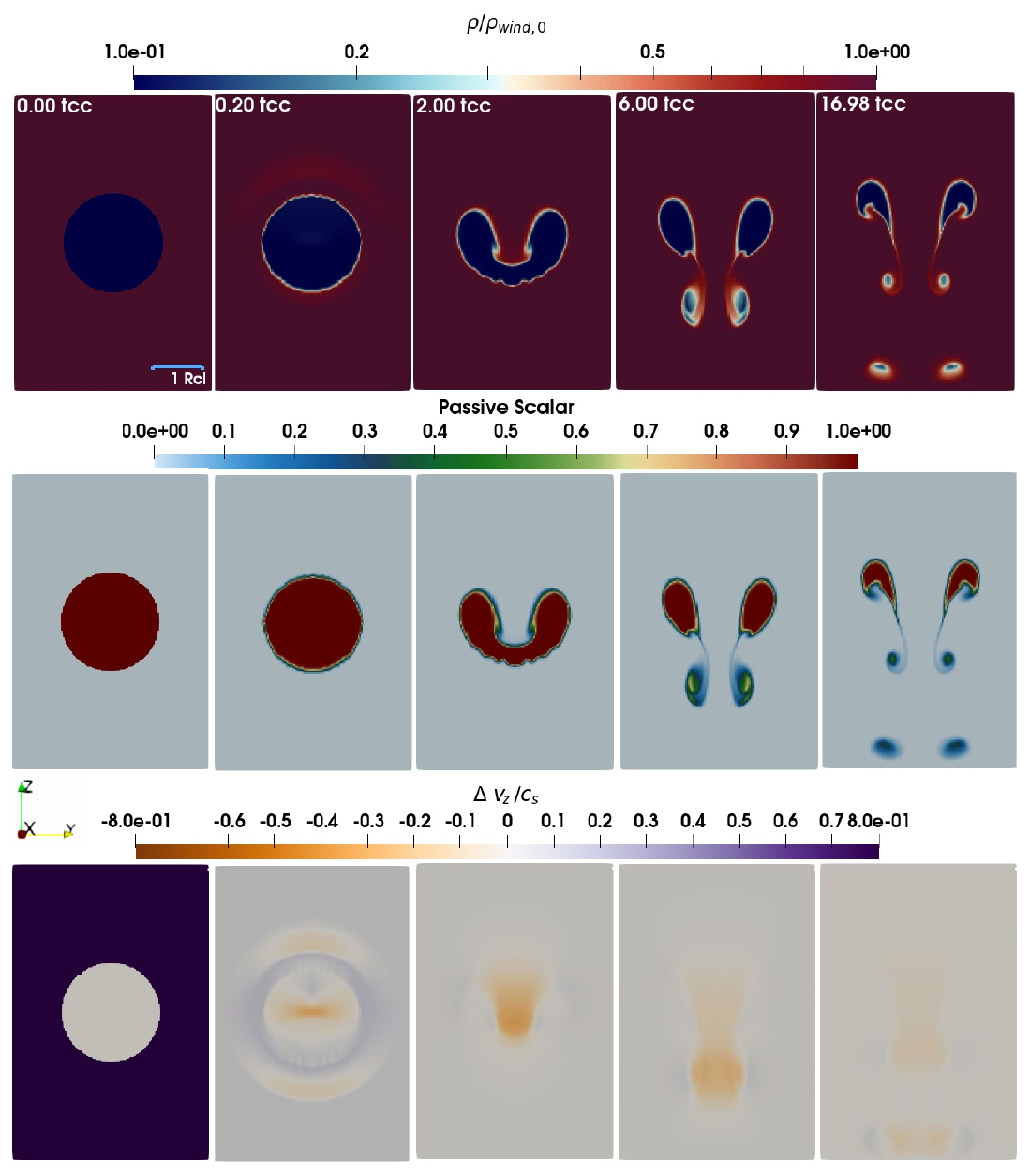}
            \caption{
            Slices through the 3D simulations of the bubble drag problem from our suite of idealized simulations of cloud-wind interactions. The bubble starts from zero velocity with a wind blowing on it all over the simulation domain from bottom to top. A zoom-in around the bubble is shown in all panels. \textit{Top row}: Density in units of initial wind density ($\rho_{{\rm wind},0}$) on a logarithmic scale. \textit{Middle row}: Passive scalar marking the bubble. \textit{Bottom row}: Relative velocity between the bubble and wind, normalized by the sound speed of the wind $c_{s}$. For an initial density contrast $\chi=0.1$ and Mach number $\mathcal{M}\equiv v_{\rm wind}/c_s = 0.8$, the bubble quickly becomes co-moving ($\sim0.2t_{\rm cc}\approx 1.4t_{\rm drag}$) with the wind and survives indefinitely, as seen in the rightmost column in density and tracer fields. Timescales in the bubble drag problem \adchange{provide an estimate of the} CGM stripping timescale, as discussed in Section \ref{subsec:TimeScaleBubble} and Figure \ref{fig:nG_comp}.
            The evolution can be seen in this video link: \href[pdfnewwindow=true]{https://youtu.be/Ulhk6ko2bMc}{https://youtu.be/Ulhk6ko2bMc}. 
            }
            \label{fig:bubble_slice}
        \end{figure*}
        In all these control simulations, the evolution of the ISM is distinctly different when compared to RPS. 
        This is expected since the evolution of the dense ISM in RPS simulations is dominated by strong gravitational attraction from the stellar disk and the dark matter halo. Minor differences in the late time evolution of the CGM mass between RPS and cloud crushing simulations in Figure \ref{fig:nG_comp} are also expected because of weak but non-zero gravity. The loss of the CGM occurs at a slower rate at late times in the 
        simulations with gravity, as a small amount of CGM always lingers in the shadow of the dense ISM where gravity is strong \rfchange{(see the CGM tracer field in Figure \ref{fig:rps_snapshots})}.
        \subsubsection{The Bubble Drag problem (Control Setup II)}\label{subsec:TimeScaleBubble}
        As outlined in the preceding section, CGM stripping and cloud crushing have common underlying physics. In this section, we analyze these physical similarities.
        
        Figure \ref{fig:bubble_slice} shows the evolution of a bubble in a dense wind. This physically interesting regime has not been explored much in the astrophysical context.
        The three rows show the evolution of density, passive scalar (marking the bubble), and relative velocity of the bubble with $\chi=0.1$ moving in a subsonic wind of Mach number $\mathcal{M}\equiv v_{\rm wind}/c_{s} =0.8$. \rfchange{As discussed in Section \ref{subsec:control_setupCGM_only}, } shear instabilities like Kelvin-Helmholtz and Rayleigh-Taylor can disrupt a cloud/bubble over a cloud crushing time $t_{\rm cc}=\sqrt{\chi_{\rm eff}} R_{\rm cl}/v_{\rm wind}$, where $\chi_{\rm eff}$ is the ratio of densities of the denser to rarer medium ($\chi_{\rm eff}=\rho_{\rm wind}/\rho_{b}$ for the bubble drag problem). 
        We see that the bubble evolves to become co-moving with the wind within $0.2\ t_{\rm cc}$ (for $\chi=0.1$ and Mach $\mathcal{M}=0.8$) and survives indefinitely. On the other hand, the time required for a dense cloud to become co-moving in the wind is longer ($t_{\rm drag, cloud}\sim \chi R_{\rm cl}/v_{\rm wind}$) before which it is completely mixed in the wind and destroyed (\citealt{Klein94, GronkeOh2018MNRAS.480L.111G, Kanjilal2021MNRAS.501.1143K}).
        
        The motion of a bubble in a dense wind can be understood by taking into account the dominant forces  (see Table 2 and Eq. 2 from \citealt{DARMANA20091077}) acting on it: (i) the \textit{drag force} resisting the motion due to relative velocity, and (ii) the \textit{virtual added mass force} resisting its acceleration through a dense medium.
        The virtual added mass is important for a bubble accelerating through a dense fluid, as to move, the bubble has to displace an additional mass of the fluid surrounding it, comparable to its own volume times the ambient density (\citealt{lambHydro, batchelor_2000}). 
        While this force would also arise for a dense cloud in a diffuse wind (cloud crushing), the added mass in that case is negligible compared to the mass of the cloud.
        The equation of motion for the bubble can be written as follows,
        \begin{equation}
            \begin{aligned}
            \frac{4\pi \rho_{b} R_{b}^3}{3}\frac{dv_b}{dt} = &-\frac{C_D \pi}{2}\rho_{\rm wind} R_{b}^2(v_{\rm wind}-v_{b})^2\\ &- \frac{C_{\rm vam} 4\pi R_{b}^3}{3} \rho_{\rm wind}\frac{dv_{b}}{dt},
            \end{aligned}
            \label{eq:bubble_em}
        \end{equation}
        where $\rho_b$ and $R_b$ are the density and size of the bubble, \rfchange{$v_b$ is the velocity of the bubble}, $\rho_{\rm wind}$ and $v_{\rm wind}$ are the density and velocity of the wind, $C_{D}$ is the drag coefficient, $C_{\rm vam}$ is the virtual added mass coefficient which depends only on the geometry/shape of the lighter fluid. The first term in the right-hand side of Eq. \ref{eq:bubble_em} is due to drag and the second term is due to the virtual mass addition. 
        
        The virtual mass term comes from the necessity to account for an increase in inertia 
        due to 
        motion between a lighter 
        bubble in a denser wind. This effect was elucidated in a seminal paper by Stokes in 1850 titled ``The effect of the internal friction of fluids on the motion of a pendulum" (\citealt{Stokes1851}; see \citealt{Krishnaswami2020, MAHMOUDI2022119} for recent reviews). The expression for the virtual added mass suggests that the effective mass of the bubble would 
        increase 
        approximately by a factor $\rho_{\rm wind}/\rho_b$, raising the effective density contrast to unity.
        Rearranging the terms in Eq. \ref{eq:bubble_em} gives
        \[ \frac{4\pi}{3} \left(\rho_{b}+C_{\rm vam}\rho_{\rm wind}\right) R_{b}^3\frac{dv_b}{dt} = -\frac{1}{2}\rho_{\rm wind}C_D \pi R_{b}^2(v_{\rm wind}-v_{b})^2\ .\]
        \rfchange{
        This expression can be further rearranged and integrated in the following form
        \[
        \centering
        \bigintss_{\ 0}^{v_{b}(t)/v_{\rm wind}} 
        \frac{d\left(v'_b/v_{\rm wind}\right)}{\left( 1 - v'_b/v_{\rm wind} \right)^2
        }
        = -  \frac{3C_D^{}} {8 R_b^{}}
        \bigintss_{\ 0}^{t}\frac{dt'} {\left(\rho_b^{ }/\rho_{\rm wind} + C_{\rm vam}\right)} 
        \]
        \begin{equation}
            \nonumber
            \left. \text{which implies,\; }\frac{v_b (t)}{v_{\rm wind}} =
            \frac{t/t_{\rm drag}}{\left( 1+ t/t_{\rm drag}\right)}\right. , 
        \end{equation}
        \begin{table}
            \centering
            \caption{Parameter space for Control Setup II: a uniform CGM (/blob) in ICM (/wind)}
            \begin{adjustbox}{width=0.97\columnwidth,center}
                \begin{tabular}{cc|cc}
                \hline
                \multicolumn{2}{c|}{
                \begin{tabular}[c]{@{}c@{}}Bubble drag \\ {[} Domain= (8,8,40) $R_{\rm cl}$ {]} \end{tabular}}
                & \multicolumn{2}{c}{
                \begin{tabular}[c]{@{}c@{}}Cloud crushing \\ {[} Domain= (10,10,100) $R_{\rm cl}$ {]}\end{tabular}} \\ 
                \hline
                $\chi = \rho_b/ \rho_{\rm wind}$   & $\mathcal{M}=v_{\rm rel}/c_{\rm s, wind}$ & $\chi = \rho_{\rm cl}/ \rho_{\rm wind}$     & $\mathcal{M}=v_{\rm rel}/c_{\rm s, wind}$  \\
                \hline
                \hline
                0.1   & 0.8  & 10  & 0.8\\
                0.1   & 0.5  & 10  & 1.2\\
                0.1   & 1.2  & 100 & 0.8\\
                0.25  & 0.5  & 100 & 1.2\\
                0.25  & 0.8  &     & \\
                0.25  & 1.2  &     & \\
                \hline
                \hline
                \end{tabular}
            \end{adjustbox}
            \label{tab:bubble drag params}
        \end{table}
        \begin{equation}
            \begin{aligned}[b]
                \text{where,\; }t_{\rm drag}={}&\frac{8}{3C_D}\left(C_{\rm vam} +\frac{\rho_{b}}{\rho_{\rm wind}} \right) \frac{R_b}{v_{\rm wind}}& \\
                \approx {}&  
                C_{\rm vam}\left( \frac{R_b}{ v_{\rm wind}} \right),
                \hspace{3em} \text{ for } \frac{\rho_b}{\rho_{\rm wind}}\ll 1\;
            \label{eq:tdrag}
            \end{aligned}
        \end{equation}
        where we have used $C_D = 8/3$. As $t/t_{\rm drag} \gg 1$, $v_{b} \rightarrow v_{\rm wind}$. Eq. \ref{eq:tdrag}} suggests that the characteristic time for the bubble drag problem is the drag time, which is independent of $\chi$.
        \adchange{Note that the above expression reduces to $t_{\rm drag}=\chi R_{b}/v_{\rm wind}$ for $\rho_b/ \rho_{\rm wind} \gg 1$, which is the usual expression of drag time in the classic cloud crushing problem ($\rho_b/ \rho_{\rm wind} \equiv \rho_{\rm cl}/ \rho_{\rm wind}$).}
        \begin{figure}     
            \includegraphics[width=\columnwidth]{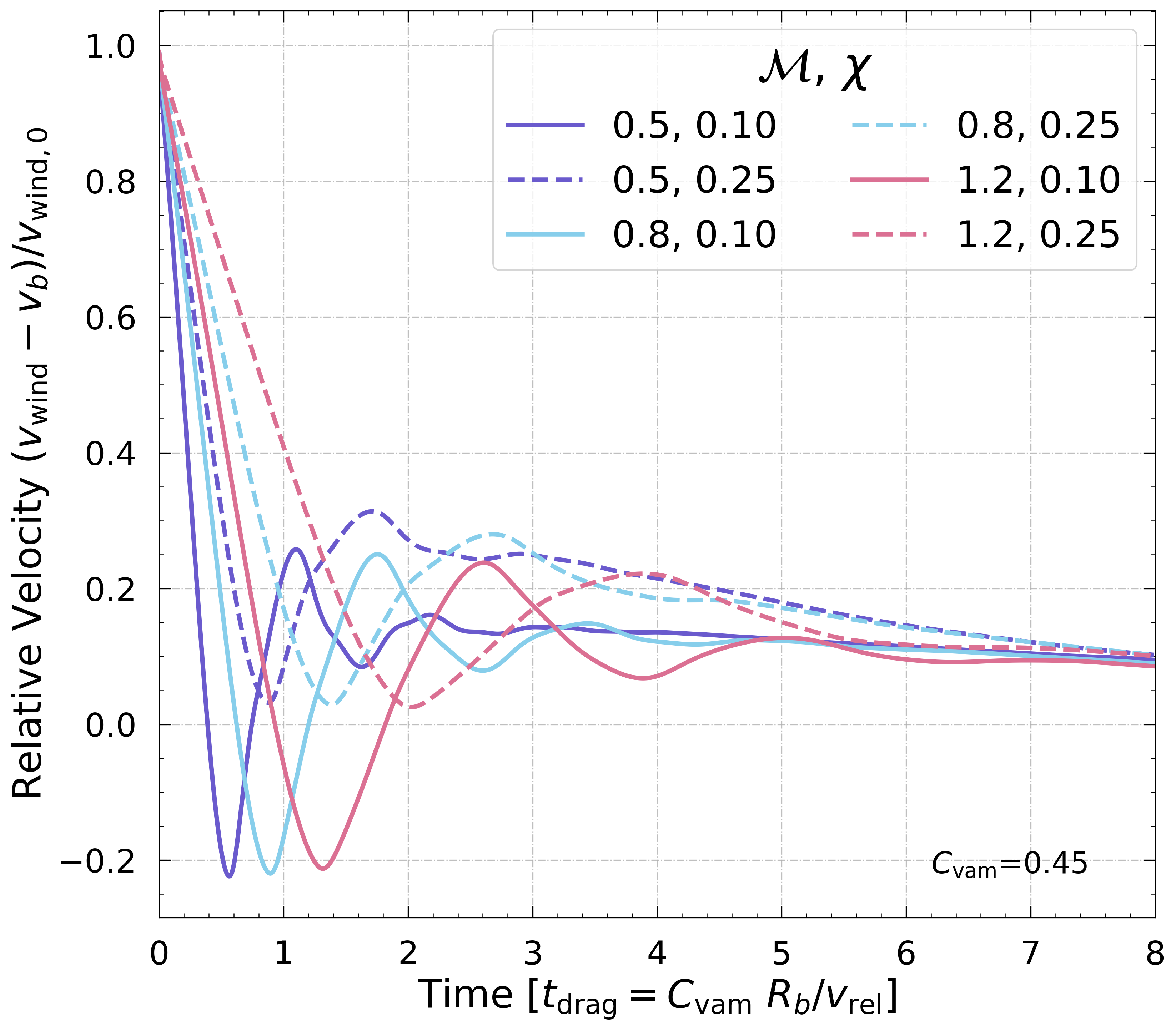}
            \caption{
            Evolution of the relative velocity between the diffuse bubble and the dense wind (normalized by the initial wind velocity $v_{{\rm wind},0}$) in the bubble drag problem.
            The various combinations of density contrast $\chi$ and Mach number of the wind $\mathcal{M}$ shown here are from our suite of idealized simulations of \textit{`cloud-wind interaction'}.
            In all the cases, the bubble becomes co-moving in about a few $t_{\rm drag}$, the estimate of which is presented in Section \ref{subsec:TimeScaleBubble}. The \textit{cyan solid} line 
            corresponds to the run shown in Figure \ref{fig:bubble_slice}. The lower $\chi$ simulations have initial signatures of high-frequency sound waves traveling inside the bubble, which are damped over a drag time. Since our parameter space of interest is $\chi \gtrsim 1/20$, we do not investigate this further.
            }
            \label{fig:bubble_velocity}
        \end{figure}
        \begin{figure}
            \includegraphics[width=0.95\columnwidth]{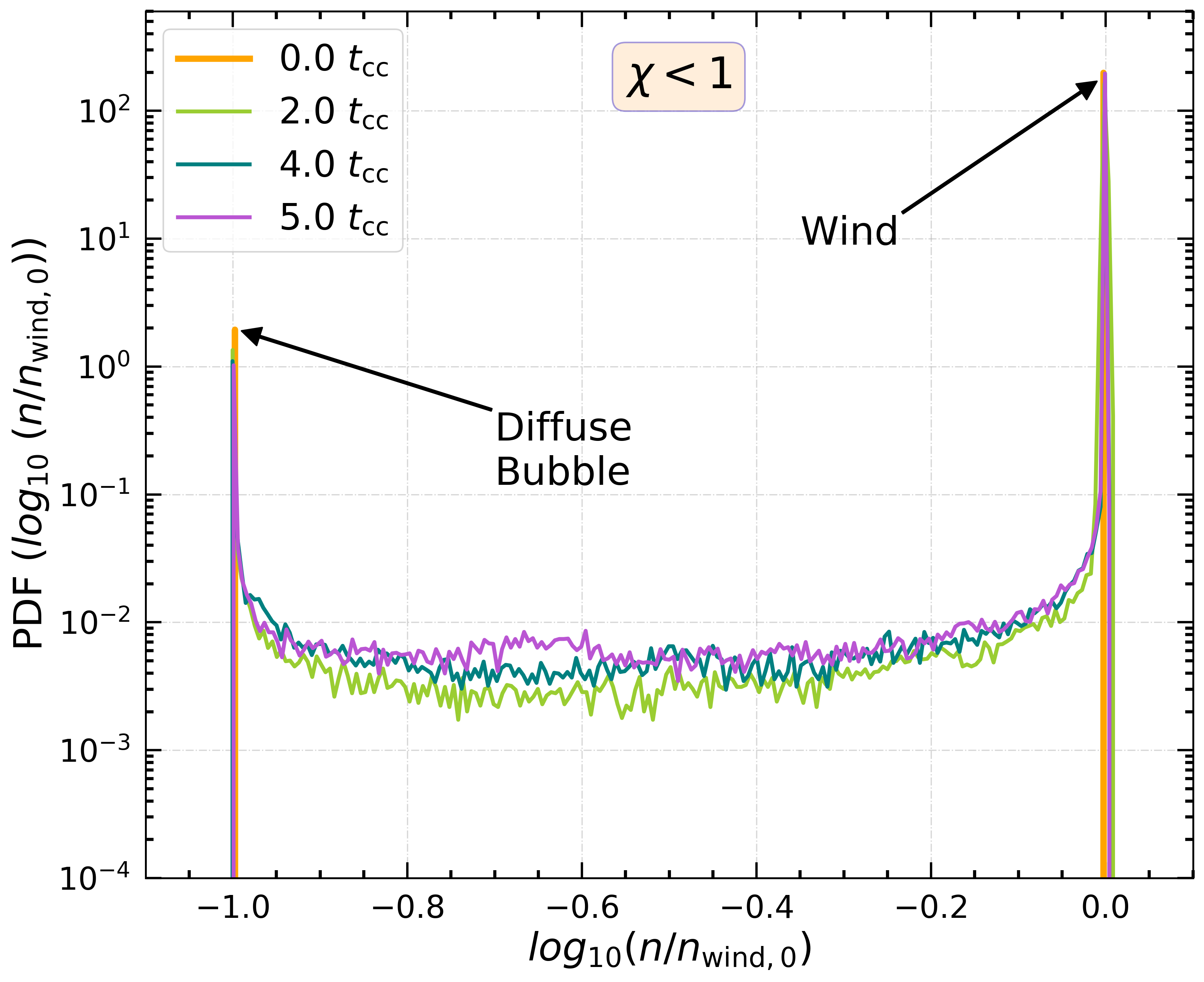}
            \vspace{0.2em}\\
            \includegraphics[width=\columnwidth]{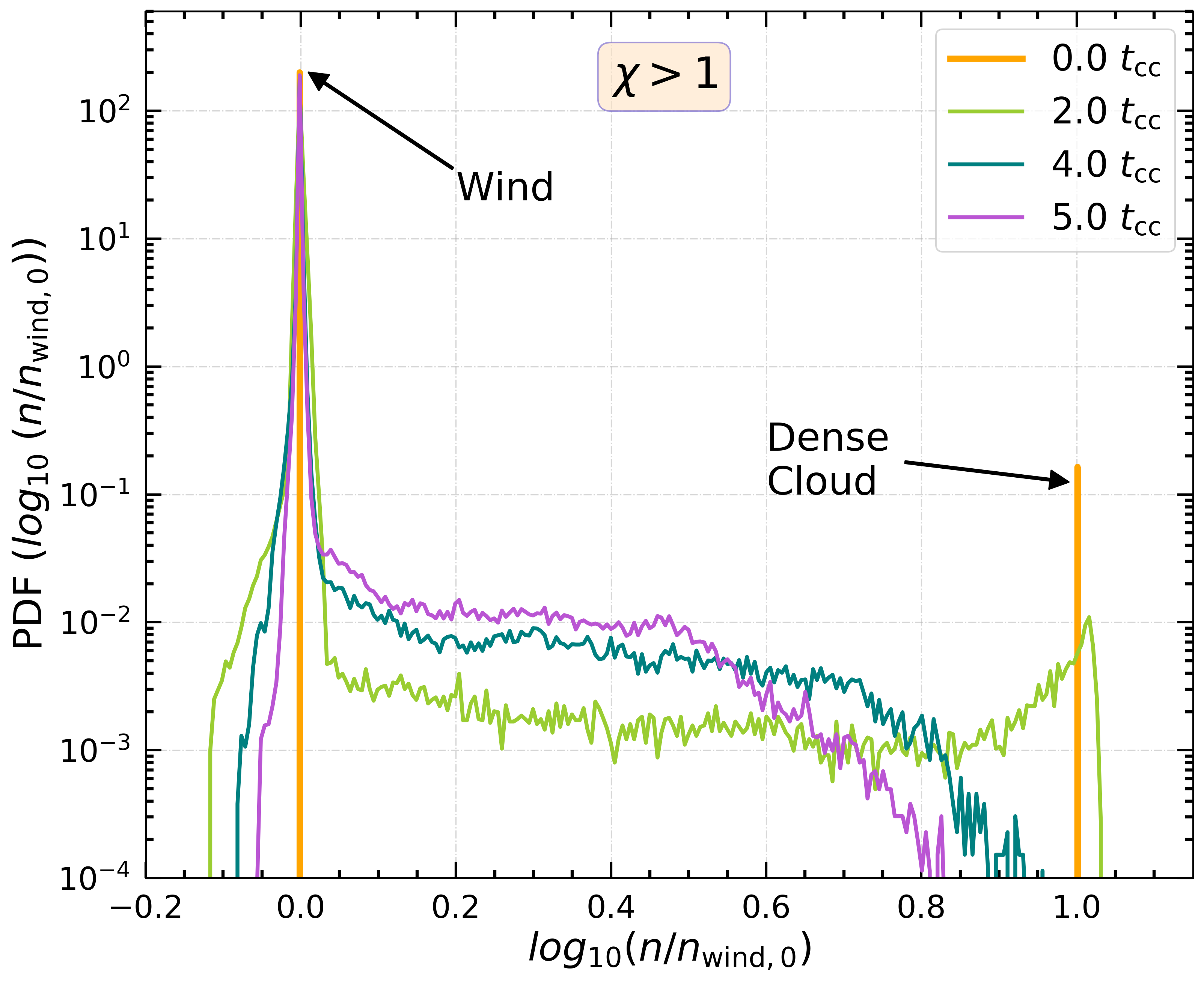}
            \caption{Volume PDF of number density at different cloud crushing times from our idealized cloud-wind interaction simulations. \textit{\rfchange{Top} panel}: The bubble drag problem ($\chi<1$). \textit{\rfchange{Bottom} panel}: The cloud crushing problem ($\chi>1$).  
            In the bubble drag problem, initial mixing causes a slight decrease and an accompanying spread in the bubble's peak. As the bubble quickly becomes co-moving (see Figure \ref{fig:bubble_velocity} and Section \ref{subsec:TimeScaleBubble} for details on timescales), shear instabilities do not grow, and the peak at the bubble's density survives indefinitely. This is in stark contrast to the cloud crushing problem, where the peak at the cloud's density is diminished in about a few cloud crushing times ($t_{\rm cc}$) due to continuous mixing by shear-driven instabilities.
            }
            \label{fig:bubble-volume-pdf}
        \end{figure}
        
        \adchange{Figure \ref{fig:bubble_velocity} shows the evolution of relative velocity between the bubble and the wind, normalized to the initial wind velocity $v_{\rm wind,0}$ (see Eq. \ref{eq:vel_COM} for definition of cloud/bubble velocity). The time is expressed in units \adchange{of} $t_{\rm drag}$, where we use $C_{\rm vam}=0.45$ in the expression obtained in Eq. \ref{eq:tdrag}.} The chosen values of the drag coefficient $C_D$ and the virtual added mass coefficient $C_{\rm vam}$ are consistent with the range expected for a spherical bubble (\citealt{wijngaarden_jeffrey_1976, MAO2022210}). The evolution at various Mach number $\mathcal{M}$ (different colors) and density contrast $\chi$ (different linestyles) is similar
        and the (normalized) relative velocity saturates to the same value ($\sim 0.1$) within a few drag time.
        As expected from Eq. \ref{eq:tdrag}, the time evolution, when normalized by $t_{\rm drag}$, is found to be independent of the density contrast $\chi$ and Mach number $\mathcal{M}$ at late times. 
        The initial acceleration of the bubble gives rise to the development of Rayleigh-Taylor instability in the head of the bubble, where the denser wind is accelerating into the diffuse bubble (visible in the density and tracer slices at $2t_{\rm cc}$ in Figure \ref{fig:bubble_slice}). Since the bubble quickly becomes co-moving (bottom row of Figure \ref{fig:bubble_slice}), the shear and acceleration go away, and the instability cannot grow further. As a result of this, the bubble only splits up, the density contrast is maintained, and it is never mixed or destroyed in the wind. Hence, the name \textit{`bubble drag'} is justified.
        
        \adchange{Figure \ref{fig:bubble-volume-pdf} shows the evolution of the volume PDFs for -- (i) $\chi<1$ (bubble drag) and (ii) $\chi>1$ (cloud crushing) problems. Comparison of the PDF at late times (violet curves at $\sim 5 t_{\rm cc}$) among the two panels demonstrates the survival of the bubble (bubble drag; top panel) in contrast to the destruction of the cloud (cloud crushing; bottom panel). Note that the density peak for $\chi<1$ survives forever (albeit with a small broadening due to initial mixing), in line with the discussion above. On the contrary, the dense cloud peak in the PDF for $\chi>1$ gradually \adchange{vanishes due to continuous} shear-driven mixing.} 
        
        \adchange{We now utilize the analytic estimates derived at the beginning of this section to obtain insights on the rate of loss in CGM mass in our simulations.}
        \adchange{Using Eq. \ref{eq:mass_loss_rate} and a virtual added mass coefficient $C_{\rm vam} = 0.45$ (appropriate for a spherical bubble; also used in Figure \ref{fig:bubble_velocity}), we plot the blue band in Figure \ref{fig:nG_comp}. This band has a slope that corresponds to $t_{\rm drag}=C_{\rm vam} R_{\rm CGM}/v_{\rm wind}$. The slope, being independent of the density contrast $\chi$ in the bubble drag ($\chi<1$) regime, accurately explains the CGM mass evolution in all the simulations corresponding to $\chi < 1$. For cases with $\chi \gtrsim 1$, corresponding to the cloud crushing regime, Eq. \ref{eq:tdrag} predicts that the mass loss rate depends on the density contrast. The two yellow bands in Figure \ref{fig:nG_comp} have slopes that correspond to $t_{\rm drag}=\chi R_{\rm CGM}/v_{\rm wind}$ and well explains the CGM mass evolution for the numerical simulations in this regime.
        }
    
    Overall, the results from our idealized simulations of \textit{cloud-wind interaction} \adchange{(the control setups I and II) highlight that -- (i) gravitational force plays a negligible role in stripping most of the CGM, and (ii) the CGM to ICM density contrast sets the stripping timescale, which can be explained by our analytic drag time estimate corresponding to either the cloud crushing (high-density contrast) or the bubble drag (low-density contrast) regime.}

\section{Discussion}\label{sec:discussion}
The primary motive for this paper \adchange{is} to investigate the hydrodynamics of \adchange{ram pressure stripping (RPS)} of the CGM and how it differs from ISM stripping. To isolate the \adchange{most important physical} effects, we ignore several \adchange{other} physical processes.
\adchange{In this section, we discuss the implications of including the CGM in the RPS of galaxies in cluster and group environments. We first address the role of CGM in communicating the ram pressure to the ISM. We then discuss the observational implications.}

    \subsection{Effect of CGM on ISM stripping}
    \adchange{Figure \ref{fig:cgm_ram_pressure} highlights the role of CGM in mediating the ISM stripping. In the top panel, solid lines show the evolution of ISM mass in the presence of a CGM for a low and a moderate ram pressure strength of the ICM wind (OG1e-3v800 [violet] and OG1e-4v800 [green], respectively). The dash-dotted lines show the corresponding evolution in the absence of a CGM. When the ICM wind directly interacts with the ISM,\footnote{For this comparative study we get rid of the CGM density and initialize the constant ICM density and velocity all over the domain, beyond the disk.} at first, we see an instantaneous stripping phase in which the outer region of the ISM is stripped (higher mass loss for higher ram pressure) in accordance with the Gunn \& Gott criterion (see Section \ref{subsec:compare_prev_work}). Subsequently, a continuous stripping phase sets in, which is driven by Kelvin-Helmholtz instabilities between the ICM and the remaining ISM, which is consistent with earlier works (cf. Figure 9 \citealt{Roediger2006}). The ram pressure faced by the disk (middle panel) is smaller for the lower ram pressure run (green dash-dotted line)  as the ISM can effectively decelerate the incoming flow of the ICM with a lower density. However, in the presence of the CGM, the onset of the instantaneous stripping phase is delayed and the ram pressure experienced is modified.
    \begin{figure}
    \centering
    	\includegraphics[width=0.98\columnwidth]{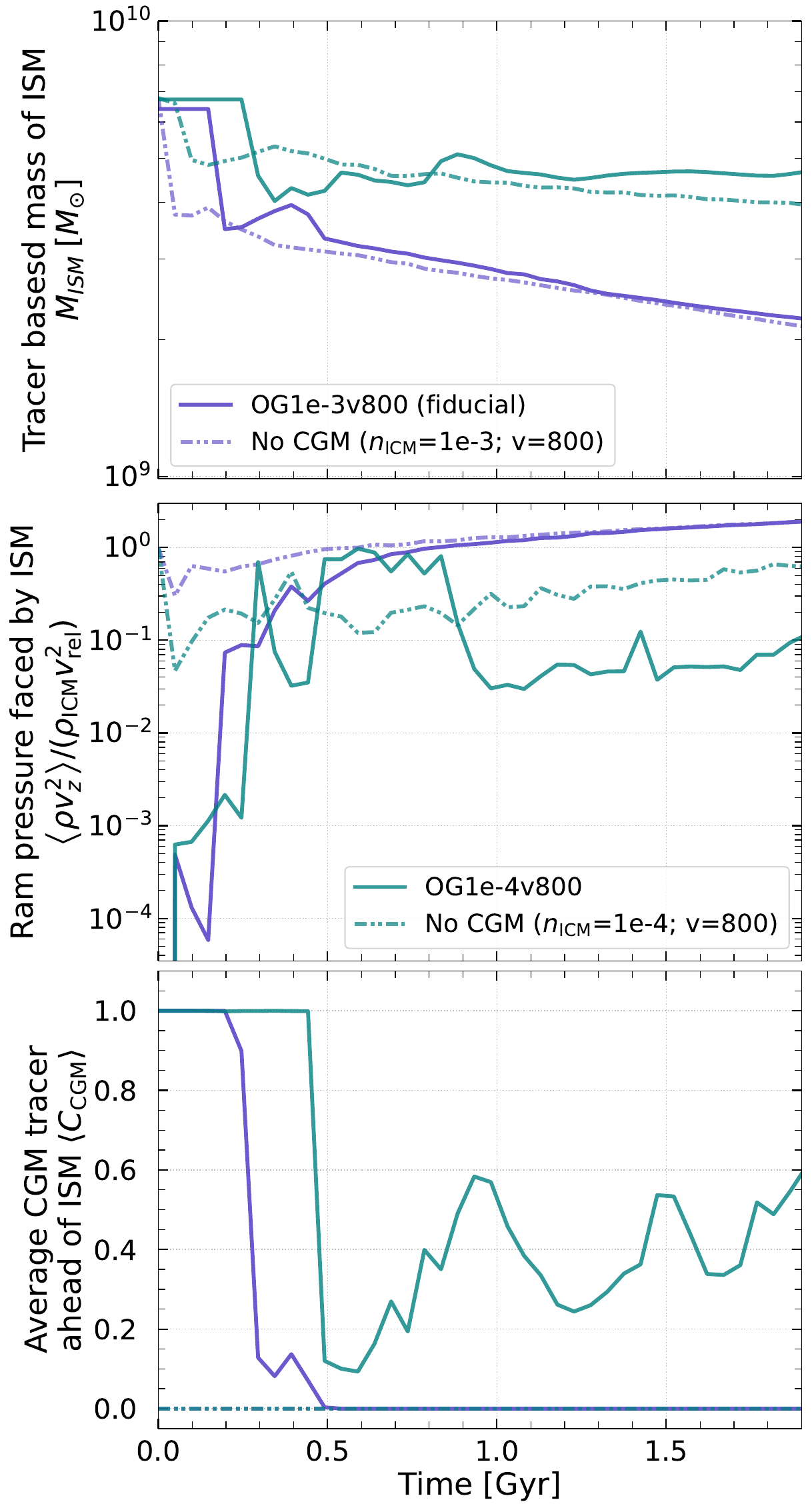}
        \caption{\adchange{
        The effect of the CGM on ram pressure faced by the ISM.
        \textit{Top panel}: The evolution of the ISM mass in the presence of the CGM (\textit{solid curves}) in comparison to those without the CGM (\textit{dash-dotted curves}).
        \textit{Middle panel}: The evolution of effective ram pressure faced by the ISM, normalized by the expected ram pressure strength of the wind ($\rho_{\rm ICM}v_{\rm rel}^2$; strength at infinity). The effective ram pressure is calculated as the mass-weighted average (denoted by $\left< ... \right >$) of the product of density and the square of z-velocity in a plane $2\ \rm kpc$ ahead of the disk, within a radius of $20\ \rm kpc$. 
        \textit{Bottom panel}: The mass-weighted average CGM tracer in the chosen plane.
        The CGM communicates a reduced ram pressure to the ISM until the CGM-ICM interface reaches close to the disk plane (when the CGM tracer falls significantly). For the low ram pressure (\textit{solid green} line), a non-negligible amount of CGM gas continues to impart reduced ram pressure on the ISM.
        }
        }
        \label{fig:cgm_ram_pressure}
    \end{figure}
    
        We analyze the effective ram pressure faced by the ISM as the mass-weighted average (denoted as $\langle \rangle$) value of $\rho v_z^2$  evaluated at a height $z=-2\ \rm kpc$ within a radial extent of $20\ \rm kpc$ (which is within the extent of bow shock that develops around the disk at late times). The evolution of the effective ram pressure $\langle \rho v_z^2 \rangle$ with respect to the imposed ICM ram pressure $\rho_{\rm ICM}v_{\rm rel}^2$ is shown in the middle panel. The bottom panel shows the mass-weighted average concentration of CGM tracer $\langle C_{\rm CGM} \rangle$ faced by the ISM. 
        
        For a moderate ICM ram pressure (OG1e-3v800 shown in the \textit{solid violet} line), the presence of CGM delays the ISM stripping. The effective ram pressure felt by the ISM is less than 10\% of the imposed $P_{\rm ram}$ due to the presence of the CGM for about $200\ \rm Myr$. 
        As the CGM density ($\sim 10^{-4}\ \rm cm^{-3}$) is lower than that of the impinging ICM ($10^{-3}\ \rm cm^{-3}$), the ISM initially faces a lower density as well as a lower ram pressure.
        However, as the CGM gets stripped away and CGM tracer $\langle C_{\rm CGM} \rangle$ drops to zero, the ICM eventually impacts the ISM directly. Beyond this, the ram pressure gradually increases to the imposed ICM ram pressure strength at infinity. The timescale at which the CGM-ICM interface reaches the disk depends on the velocity of the forward shock induced by the ram pressure. Estimating this time is complicated due to the acceleration of the wind because of gravity and is beyond the scope of the present work.
        
        In contrast, at lower ICM ram pressure (OG1e-4v800 shown in the \textit{solid green} line), the CGM provides more effective shielding, resulting in less drastic ISM stripping over longer timescales.
        Over $400\ \rm Myr$, the CGM shields the ISM completely. After that, the CGM concentration begins to drop, and the ram pressure on the ISM is enhanced, which now corresponds to the instantaneous stripping phase. A sudden compression, seen in this case, may lead to a momentary increase in star formation and AGN activity (e.g., \citealt{Poggianti2017, Akerman2023}) near the disk of ram pressure stripped galaxy, but will ultimately lead to enhanced stripping of the cold ISM and eventual quenching of star formation. Moreover, at late times, some amount of CGM always lies ahead of the disk (non-zero $\langle C_{\rm CGM} \rangle$) as restoring pressure due to gravity becomes non-negligible. The CGM-ICM interface\footnote{The reduced $\langle C_{\rm CGM} \rangle$ is mostly compensated by an increase in $\langle C_{\rm ICM} \rangle$ and a small fraction of $\langle C_{\rm ISM} \rangle$.} progressively pushes the remaining gas out of the ISM, but at a lower ram pressure strength. The ISM mass in the continuous stripping phase is, therefore, slightly higher compared to the case without CGM. 

        }
        
    \subsection{Caveats}
    \adchange{One of the limitations of our model is the assumption of treating the CGM and ISM as isothermal components. Observational evidence suggests that both the ISM and the CGM are multiphase (\citealt{XavierProchaska2013, Tumlinson/annurev-astro-091916-055240, Chen2018, Zahedy2018}). Although ram pressure stripping depends on the combination of density and relative velocity, accounting for the range of temperatures as well as metallicities in the ISM and CGM would be crucial in determining the neutral and ionized gas fractions in the stripped tails. 
    }
    
    An important physical process not included in our work is radiative cooling. However, the timescale estimates for ISM+CGM stripping are likely to remain unaffected, as the volume-filling diffuse CGM has a very low density and a long cooling time. The stripped ISM gas near the ISM-CGM/ICM boundary will definitely cool, may grow because of its short cooling time, and show a different morphology. However, hydrodynamic mixing and an increased fraction of ICM/CGM down the tail are expected to be qualitatively similar. \adchange{
    The sizes of 
    dense clouds produced by radiative cooling can 
    exceed} the Jeans length and become gravitationally unstable, leading to in-situ star formation (e.g., \citealt{Vulcani2018,Cramer2019}). Other \adchange{important} physical processes that we do not model \adchange{are the effects of} magnetic fields, thermal conduction, and viscosity. \adchange{These can lead to suppression of mixing of the clouds in the stripped tails}, potentially affecting the morphology of dense cold gas (\citealt{Ruszkowski2014, Tonnesen2014,Mueller2021}), accounting for which is beyond the scope of our current work. 
    
        \subsubsection{\adchange{CGM and ICM density contrast}}
        \adchange{
        If a moderately massive satellite, like our model galaxy with $M_{\rm 200}=10^{12} \ M_{\odot}$, starts 
        near the cluster virial radius, the initial density contrast between the CGM and the ICM would have $\chi_0 > 1$. 
        For example, with an average CGM density of $10^{-4} \ \rm cm^{-3}$ in our model galaxy, the initial density contrast $\chi_0\sim 10$, if it is at distance of $1.5\ \rm Mpc$ (ambient density $\sim 10^{-5}\ \rm cm^{-3}$) from a Virgo-like cluster core. 
        As this galaxy falls into the denser 
        core, its instantaneous density contrast $\chi$ will vary as the ambient medium becomes progressively over-pressurized. Assuming a one-zone model for the CGM, with an instantaneous size $R_{\rm CGM}$ and an instantaneous density contrast $\chi$ surrounded by an ambient ICM of density $\rho_{\rm ICM}$, two important timescales emerge. First, on a dynamical time  $t_{\rm dyn}=r_{\rm ICM}/v_{\rm wind}$, the galaxy would fall towards the cluster core (assuming $r_{\rm ICM}$ as the virial radius for the cluster). Second, over a collapse time $t_{\rm collapse}\sim R_{\rm CGM} / \sqrt{\gamma P_{\rm ICM}/\rho_{\rm CGM}} \sim \chi^{1/2} R_{\rm CGM} / c_{\rm s, ICM}$, the CGM pressure would achieve pressure equilibrium with the local ICM. If $t_{\rm collapse}> t_{\rm dyn}$ and the ICM density increases from cluster outskirts $\rho_{\rm ICM,0}$ to the core by a factor $\sim 10^3$, the infalling galaxy will move isochorically and would have a density contrast $\chi = \chi_0 \rho_{\rm ICM,0}/\rho_{\rm ICM}\sim 0.01$. Therefore, if the dynamical time is short, the CGM would evolve isochorically, leading to a density contrast $\ll 1$, which falls deep into the bubble drag regime.
        
        In contrast, if $t_{\rm collapse}< t_{\rm dyn}$, the effective density contrast would not be as low as the 
        isochoric case.
        If we 
        assume that the CGM evolves adiabatically, or the entropy 
        increases (due to the shock moving into the CGM) by a factor $f_>$ (greater than unity), such that
        \[
        \frac{P_{\rm CGM}}{{\rho_{\rm CGM}}^{\gamma}} = f_>\ \frac{P_{\rm CGM,0}}{{\rho_{\rm CGM,0}}^{\gamma}}.
        \]
        If the CGM is able to achieve pressure balance with the ambient ICM at its instantaneous position,
        \[
        P_{\rm CGM} = f_P \left(P_{\rm ICM} + P_{\rm ram}\right),
        \]
        where $P_{\rm ram}=\rho_{\rm ICM}v_{\rm rel}^2$ is the instantaneous ram pressure faced by the CGM, $P_{\rm ICM}$ is the ambient thermal pressure and $f_P$ is 
        a pressure-factor ($\lesssim 1$) to account for partial pressure equilibrium.
        Rearranging the above equation, we find that
        \begin{equation}
        \nonumber
            \left[  
                \frac{f_P}{f_{>}}\frac{\rho_{\rm ICM}\left( {c_{\rm s, ICM}}^2 + {v_{\rm rel}}^2\right)}{\rho_{\rm ICM,0} \chi_0\ {c_{\rm s, CGM,0}}^2} \right]^{1/\gamma}=
                \frac{\rho_{\rm CGM}}{\rho_{\rm CGM,0}}
                 = \frac{\chi}{\chi_0}\frac{\rho_{\rm ICM}}{\rho_{\rm ICM,0}},
        \end{equation}
        where $c_{s, \rm CGM/ICM}$ is the sound speed in the CGM/ICM.
        This suggests that the effective density contrast of the CGM at a later time would be
        \begin{equation}
            \chi = {\chi_0}^{(\gamma-1)/\gamma}  
                    \left( 
                         \frac{\rho_{\rm ICM}}{\rho_{\rm ICM,0}}
                    \right)^{-(\gamma-1)/\gamma}
                    \left[ 
                        \frac{f_P}{f_{>}}
                        \left(\frac{c^2_{\rm s, ICM} + v^2_{\rm rel}}{{c^2_{\rm s, CGM,0}}}
                        \right)
                    \right]^{1/\gamma}.
        \end{equation}
        Therefore, for an increase in ambient ICM density from cluster outskirts to the core of $\rho_{\rm ICM}/\rho_{\rm ICM,0}\sim 10^3$, and an initial density contrast $\chi_0=10$, the density contrast at late times, $\chi \approx (10^{-0.8}\times (10(\mathcal{M}^2+1)/6)^{0.6})\sim 0.5$, assuming $f_P=1$ and $f_{>}=6$ (from our entropy snapshots), a Mach number of 2. Here we assume the ICM to be isothermal with $c^2_{\rm s, ICM}/c^2_{\rm s, CGM,0} = 10^7 {\rm~K}/ 10^6 {\rm~K}$. 
        
        The true $\chi$ as the galaxy moves to the cluster core will lie between our isochoric and isobaric limits above, depending on the ratio of the dynamical time and the collapse time. For typical parameters, the CGM in the cluster core is expected to be in the bubble drag regime, but not with $\chi \ll 1$ if it is able to achieve pressure equilibrium with the surrounding ICM. A corollary of the above is that the galaxies that do not go close to the cluster core are always in the cloud crushing regime.
        }

    \subsection{Astrophysical Implications}
        \subsubsection{\adchange{CGM stripping in clusters}}
        \adchange{
        Massive galaxies in clusters host a substantial circumgalactic reservoir of gas. The stripping rate of this CGM is dependent on the galaxy's location within the host cluster and its history (e.g., CGM may be lost for good after a pericentric passage close to the cluster core). Galaxies farther from the cluster center experience a more gradual removal of their CGM, whereas those orbiting near the dense central regions undergo a rapid CGM stripping. The `bubble drag regime' identified in our work 
        provides a qualitative understanding of CGM removal when the ambient density is 
        high. Such a scenario can be observed in the Zwicky cluster Z8338, where  \citealt{Schellenberger2015A&A} report a luminous X-ray tail with thermal emission at a temperature of $0.77\rm \ keV$ (lower than the cluster temperature $\sim 3\ \rm keV$). This X-ray tail with luminosity of $2 \times 10^{42}\rm \ erg\ s^{-1}$, has been found by the authors to be at the same redshift as the nearby elliptical galaxy CGCG254-021 (at a distance of $40\rm \ kpc$ from the head of the luminous tail), which has lost all of its X-ray emitting gas. We suggest that this tail can be formed from the complete stripping of the hot halo/CGM of CGCG254-021. The galaxy can be undergoing stripping in the bubble drag regime as it is at a projected distance of $310\rm \ kpc$ from the cluster center (cluster mass $\sim 10^{15}\ M_{\odot}$ and virial radius $\sim 1\rm \ Mpc$). The ambient ICM  density reported by \citealt{Schellenberger2015A&A} is $(8.0 \pm 0.3) \times 10^{-28}\rm \ g\ cm^{-3}$, which is greater than the average CGM density of such a massive galaxy ($\sim 1 - 4 \times 10^{-28}\rm \ g\ cm^{-3}$). This makes it possible for the CGM to be detached from the galaxy's potential well, as it becomes comoving with the ICM wind over a crossing time. An ambient high pressure in this region would compress the CGM-like bubble as it is getting stripped and would result in an enhanced X-ray emitting tail.
        } 
        
        \adchange{
        We looked at galaxies in the Virgo cluster ($M_{\rm 200}=1.14- 4.2 \times 10^{14}\ M_{\odot}$, $r_{\rm 200}=1.08\ \rm Mpc$; \citealt{Schindler1999, Mei2007, Urban2011}) and found 
        examples of how identifying the present stripping regime of a galaxy can elucidate its orbital history. To this end, we compare the galaxies M89 (NGC 9552) and M90 (NGC 4569). 
        \textit{Chandra} observations reveals M89's X-ray bright tail of stripped gas at $0.5\ \rm keV$ with gas densities of $(5.4\pm 1.7)\times 10^{-3}\ \rm cm^{-3}$ in the tail and $\sim 10^{-2}\ \rm cm^{-3}$ in the core (\citealt{Machacek2006}). The line-of-sight velocity of M89 is $1700 \ \rm km \ s^{-1}$ relative to the Virgo cluster center (M87) at a projected distance of 350 kpc i.e., $0.32\ r_{\rm vir}$ (given that the line-of-sight velocity is high, the actual distance of M89 can be similar to the projected distance).
        \citealt{Roediger2015a, Roediger2015b} simulated possible orbits for M89 ($M_{\rm DM}=5\times10^{12}\ M_{\odot}$, $r_{\rm 200}=200\ \rm kpc$) with a pericentric distance comparable to its present location. 
        Within this orbit, the gas in M89 is on an average denser ($n_e>10^{-4}\ \rm cm^{-3}$ at $100 \ \rm kpc$; cf. Figure 11 there) than the surrounding ICM ($n_{e, \rm ICM}\lesssim 3\times 10^{-4}\ \rm cm^{-3}$; cf. Fig. 12 of \citealt{Roediger2015a}), suggesting CGM stripping in the cloud crushing regime. M89's CGM can therefore survive at least a drag time $t_{\rm drag}= \chi R_{\rm 200}/v_{\rm rel}\sim 1.1\rm \ Gyr$ (assuming $\chi \sim 10$). 
        
        In contrast, the nearby galaxy M90 (a massive galaxy with $D_{25}$ of $47$ kpc and $M_{*}=10^{10}\ \rm M_{\odot}$), lying north-east to M89, shows a very limited X-ray emission (see Fig. 1 of \citealt{Boselli2016A&A}; X-ray mass $\sim 10^7\ \rm M_{\odot}$). It is moving in a highly eccentric orbit with a line-of-sight velocity $-235\ \rm km \ s^{-1}$ (with a major velocity component in the plane of the sky) with respect to the Virgo cluster center. Given that its stellar mass is comparable to the Milky Way, M90 can host a $\sim 200\ \rm kpc$ CGM with an average density of $10^{-4}\ \rm cm^{-3}$. \citealt{Boselli2016A&A} report an ambient ICM density of $4 \times 10^{-5}\ \rm cm^{-3}$, implying that at its present location, M90's CGM could have survived (as $\chi>1$) if it 
        is undergoing its first passage through the cluster. 
        However, the narrow-band H$\alpha+[\rm NII]$ imaging of M90's tail with \textit{MegaCam} (\citealt{Boselli2016A&A}) suggests that M90 is orbiting from west to east, and this galaxy has already encountered the maximum ICM density in its orbit at a radial distance of $\gtrsim 230\ \rm kpc$ (north to M87). At $230\ \rm kpc$, an ambient density of $8 \times 10^{-4}\ \rm cm^{-3}$ (Fig 12. of \citealt{Roediger2015a}) puts M90's CGM in the bubble drag regime. Therefore, M90 would have lost its CGM within $t_{\rm drag}=C_{\rm vam}\times R_{\rm 200}/v_{\rm rel}\sim 100 \ \rm Myr$ ($R_{\rm 200}=200\ \rm kpc$, $C_{\rm vam}=0.45$ and assuming $v_{\rm rel}$ is at least $1000\rm \ km \ s^{-1}$) in its pericentric passage.
        Therefore, the paucity of its X-ray emission, in stark contrast to the prominent X-ray tail of M89 at an even smaller cluster-centric distance, can be ascribed to M90 having previously undergone a pericentric passage close to the cluster core. 
        }
        
        \adchange{
        Another interesting observation of a galaxy undergoing stripping in the cloud crushing regime is that of M86 (NGC4406) in the Virgo cluster. 
        The extended plume of X-ray hot gas around M86 (first observed with \textit{Einstein} X-ray Observatory; \citealt{Forman1979ApJ}) can survive at its present position ($0.4 \pm 0.8 \rm \ Mpc$; \citealt{Mei2007}). A line-of-sight velocity of $1550 \rm \ km \ s^{-1}$, an ICM density of $\sim 2\times 10^{-4}\ \rm cm^{-3}$, and assuming a density contrast of 10 between the CGM and the ambient ICM (from \textit{XMM-Newton} observation; \citealt{Ehlert2013}), the characteristic timescale for its CGM stripping is estimated to be $\sim 2 \ \rm Gyr$. Such a long timescale, therefore, supports the presence of a large reservoir around M86.
        }
        
        
        \adchange{\citealt{Poggianti_2019} observe a significant amount of X-ray emission from the galaxy JW100 in the cluster A626. The authors attribute the enhanced X-ray emission to the mixing of stripped gas from the dense ISM and the hot ICM, which indeed is a plausible scenario as emission scales as the square of density and the stripped ISM being dense can contribute a lot after mixing with the ICM. 
        However, the authors rule out the possibility of the hot halo contributing to this emission as they suggest that the inner halo would be advected out in $\sim15$ Myr (see Appendix A therein). 
        CGM stripping is not instantaneous (as seen in our work, as well as in \citealt{McCarthy2007}) and their suggested timescale of $15$ Myr is even smaller than the crossing time across the halo of JW100. 
        With a halo mass of $1.3\times 10^{13}M_{\rm \odot}$, JW100 can, in fact, support a hot halo of size $R_{\rm JW100}=480\ \rm kpc$ at 
        a (virial) temperature of $5\times 10^6\rm \ K$. 
        It is at a projected distance of $83 \rm \ kpc$ from the cluster center ($\sim 0.1 r_{200}$ of A626; see Fig. 3 in \citealt{Poggianti_2019}), where the ambient ICM density ($5.8\times 10^{-27}\rm g\ cm^{-3}$) is much higher than the density of the hot halo ($\sim 10^{-28}\ \rm g\ cm^{-3}$; assuming self-similarity of JW100 with the Milky Way). This puts JW100 well within the bubble drag regime. Therefore, if JW100 has been moving with at most its present-day velocity of $1800\rm \ km \ s^{-1}$, its hot halo would be stripped over at least a drag time $t_{\rm drag}\sim R_{\rm JW100}/v_{\rm rel}\sim 265 \ \rm Myr$ ($ \gg 15 \ \rm Myr$). Therefore, the CGM contribution to the X-ray luminous tail of JW100 cannot be trivially ruled out. Enhanced X-ray luminosities are also seen around JO201 (at $\sim 0.2\rm r_{\rm 200}$; \citealt{Campitiello2021}) and JO194 (\citealt{Bartolini2022}), which are above the expected luminosities from their star formation rate, and can have contributions from the stripped CGM.
        } 
        \subsubsection{CGM stripping in lower mass halos: Magellanic Corona}
        \adchange{The ISM+CGM stripping is also important for group environments (\citealt{Hester_2006}). \citealt{Krishnarao2022Natur.609..915K} find OVI ions tracing $\sim 10^{5.5}\ \rm K$ gas within $50\ \rm kpc$ in the halo of the Large Magellanic Cloud (LMC) as well as CIV ions that trace the interfaces between these cold ($10^4$ K) clouds and the hot ($10^{5.5}$ K) corona -- indicating the ubiquity of mixed multiphase gas. These observations, therefore, reveal that the Magellanic group (comprising of Large and Small Magellanic Cloud with a truncated CGM around the LMC) is undergoing ram pressure stripping in the Milky Way. Our simulation results are qualitatively applicable to stripping in such `group environments'. 
        
        The observed heliocentric line-of-sight velocity of LMC is $262.2\ \rm km \ s^{-1}$ (\citealt{vanderMarel_2014}).
        The LMC can support a $110\rm \ kpc$  CGM with an average density of $\sim 5\times10^{-5}\ \rm cm^{-3}$ (see Figure 2 of \citealt{Lucchini2023arXiv231116221L}).
        Falling towards the Milky Way, the CGM of LMC would be denser than the ambient medium until reaching approximately $80 \ \rm kpc$ from the Galactic center. 
        Assuming that the LMC has been moving at velocities comparable to its present velocity $\sim 300 \ \rm km \ s^{-1}$ for an extended period, it has likely experienced CGM stripping in the cloud crushing regime for the past $\sim 3\ \rm Gyr$ ($t_{\rm drag}\sim \chi R_{\rm LMC, CGM}/v_{\rm rel}$) with a density contrast $\chi\sim 10$ till a Galactocentric distance of $\sim 80 \rm \ kpc$ (cf. Fig. 2 of \citealt{Lucchini2023arXiv231116221L}). However, given LMC is currently at a distance of $50\ \rm kpc$ from the center of the Milky Way, we hypothesize that its CGM stripping has transitioned to the bubble drag regime. The present ambient density of $ 10^{-4}\ \rm cm^{-3}$ (\citealt{Lucchini2023arXiv231116221L}) can strip the presently observed truncated ($50\ \rm kpc$) CGM on a timescale of at least $170\ \rm Myr$ ($t_{\rm drag}\propto R_{\rm trucated}/v_{\rm rel}\sim 50\ {\rm kpc}/300 \ \rm km \ s^{-1}$). This stripping timescale exceeding the travel time from $80$ to $50 \rm \ kpc$ ($\sim 100\ \rm Myr$) explains the survival of a fraction of the LMC's CGM in proximity to its disk.
        
        Simulations by \citealt{Lucchini2020} also support the CGM stripping scenario for the LMC where its circumgalactic reservoir can contribute $\approx\ $20\% to the ionized gas mass in the Magellanic stream. 
        The stripped tail in our fiducial simulation demonstrates the co-existence of the trailing ISM and the stripped CGM down the wind (see Figure \ref{fig:volumeRendering} \& \ref{fig:rps_snapshots}), thereby increasing the amount of ionized gas in the tail. 
        However, the Large and the Small Magellanic clouds also have a Magellanic Bridge among them as well as a Leading Arm ahead of LMC, which indicates tidal interactions. Consideration of realistic orbits, profiles for the CGMs of both the Milky Way and the LMC, and accounting for these tidal perturbations are necessary for a direct comparison with observations, which is beyond the scope of this work.
        }

        \subsubsection{Metallicity gradients along stripped tails}
        Spectroscopic observations by MUSE (\citealt{Franchetto2020}) show that the metallicity in the stripped tails decreases with increasing distance from the stripped galaxy. The interpretation of this observation is that the metal-rich stripped ISM is increasingly mixed with a higher fraction of the lower-metallicity ICM \adchange{and CGM} over $\sim 20$ kpc behind the ISM wake. This is qualitatively consistent with the simulations of \citet{Tonnesen2021ApJ...911...68T} and with even our simulations that do not include radiative cooling (see the filled circles in Fig. \ref{fig:volumeRendering} and \adchange{ISM} tracer panels in Fig. \ref{fig:rps_snapshots} at late times). Similar to the rotation \adchange{in the tail} inherited from the ISM disk seen in our simulations (see the second column in Fig. \ref{fig:rps_snapshots}), observations also show signatures of rotation in stripped tails (\citealt{Franchetto2020, Luo2023}). Further out along the tails, the ICM turbulence dominates the line-of-sight velocity (\citealt{Li2023}), but we do not model a turbulent ICM in this work. 

    \subsection{Comparison with previous simulations}\label{subsec:compare_prev_work}
    Idealized simulations of RPS often consider a direct interaction of the ICM wind with the ISM (e.g., \citealt{Roediger2006, Tonnesen&Bryan2010ApJ}). We find that including the CGM has non-trivial implications not only on the stripping timescales but also on the morphology of 
    the stripped galaxies. 
    
    \citealt{Mori_2000} found that spherical halos of dwarf galaxies can be rapidly lost if the ram pressure exceeds the thermal pressure in the galactic center. They suggest a stripping timescale of $t_0=2r_0/v_{\rm fs}$, where $r_0$ is the core radius of the dark matter potential and $v_{\rm fs}$ is the velocity of the forward shock of the impinging wind (proportional to $\sqrt{\chi}$; $\chi$ is the CGM to ICM density contrast). All of their simulations consider a density contrast $\chi$ 
    greater than unity and the dominant stripping mechanism was identified as mixing due to Kelvin-Helmholtz instability. 
    We find that the CGM in $\chi>1$ regime is stripped over a drag time $t_{\rm drag}=\chi R/v_{\rm wind}$ which is $\sqrt{\chi}$ longer than the estimate of \citealt{Mori_2000}. \adchange{Although shear instabilities can mix the CGM and ICM over a cloud crushing time ($\propto \sqrt{\chi}R/v_{\rm rel}$), unless the mixed gas leaves the virial radius, it can potentially refuel the ISM. However, a few cloud crushing times can become comparable to the drag time for large density contrasts ($\chi>1$) in our parameter space.} 
           
    \citealt{Steinhauser2016} find that star formation is quenched in disk galaxies as their hot gas halo (CGM) gets stripped. 
    \adchange{\citealt{Steinhauser2016} have primarily considered orbits with high relative velocities (>1000 $\rm km\ s^{-1}$) and the ambient ICM densities range from $10^{-28}-10^{-26}\ \rm g\ cm^{-3}$. 
    However, the position versus velocity phase space in the GASP survey  (cf. Fig. 7 of \citealt{Jaffe2018}) indicates a significant population of cluster galaxies with typical velocities close to the cluster velocity dispersion $\sigma_{\rm cl}$ ($\sim 400$ to $1300\ \rm km \ s^{-1}$). We have, therefore, studied a wider parameter space of relative velocities and density contrast, which reveals the effectiveness of stripping in the bubble drag regime at small cluster-centric radii.
    }
    \rfchange{We can understand the physical mechanism behind the rapid loss of CGM of the galaxy G1a studied by \citealt{Steinhauser2016}. In a close pericentric passage $100\ \rm kpc$ from the cluster core (orbits A \& C in their work), an ambient density of $\sim 6.4 \times 10^{-27}\ \rm g \ cm^{-3}$ and an average CGM density from the $\beta-$ profile of their galaxy, suggests $\chi \sim 0.02$. This would strip the CGM in the bubble drag regime over a few hundred Myr, which is also seen in our simulations (cf. magenta curve in the bottom panel of Figure \ref{fig:cgm_vs_ism_mass}). 
    }

    \begin{figure}
    \centering
    	\includegraphics[width=\columnwidth]{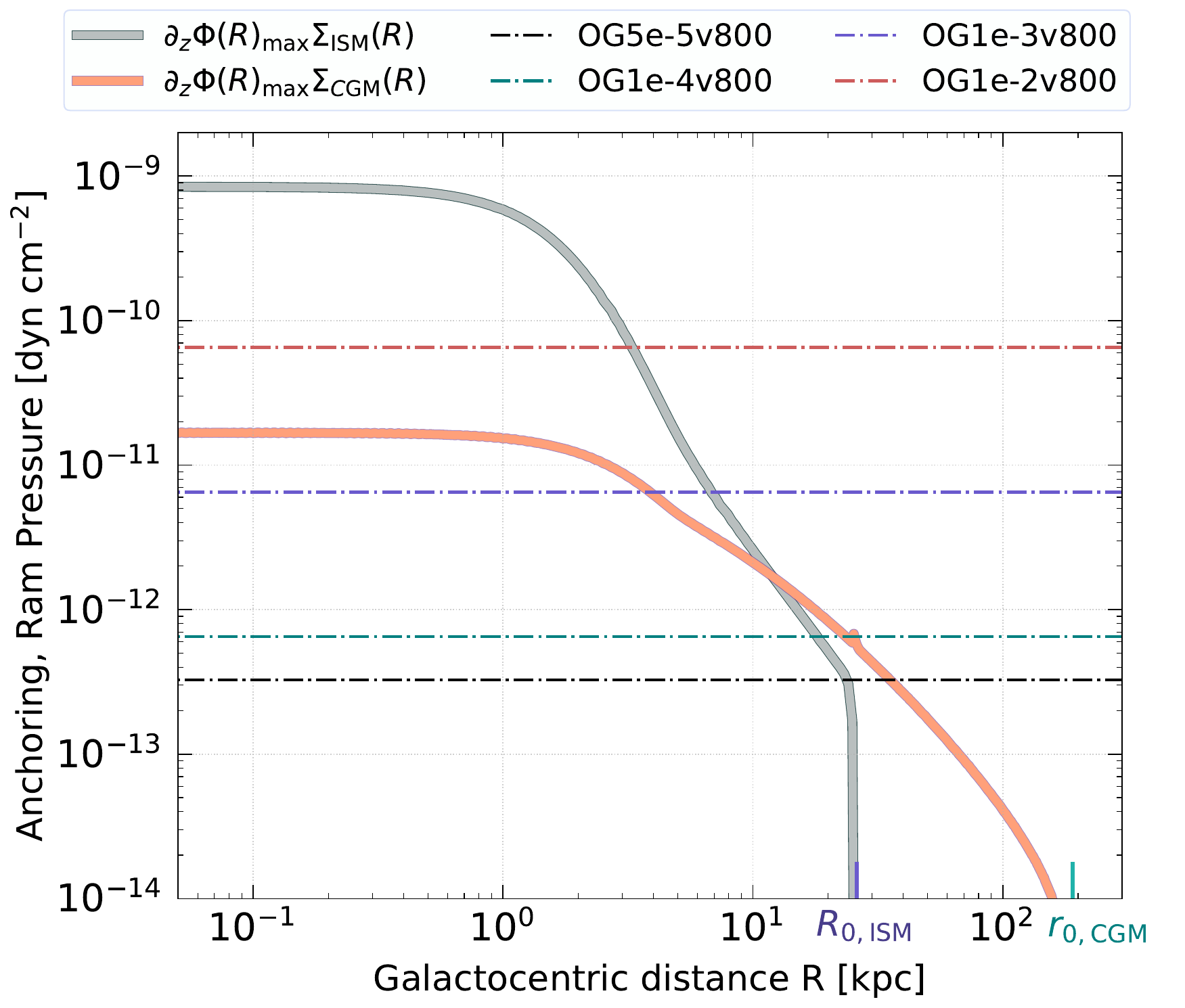}
        \caption{\adchange{The anchoring pressure for the ISM (\textit{solid gray} line) and the CGM (\textit{solid orange} line) for our model galaxy as a function of galactocentric distance $R$ (at t=0). The anchoring pressure is evaluated using the product of the maximum value of gravitational acceleration along the vertical direction $\partial \Phi/\partial z |_{\rm max}$ and the surface density of gas in the ISM/CGM, 
        $\Sigma_{\rm ISM/CGM}(R)=\int \rho_{\rm ISM/CGM} (R,z) dz$ (ISM: gas with temperature $T<3 T_{\rm ISM}$ \& CGM: $3 T_{\rm ISM} < T < T_{\rm ICM}/3$). 
        The various horizontal \textit{dash-dotted lines} represent the ram pressure strength $P_{\rm ram}=\rho v_{\rm rel}^2$ for our simulations, with the same color as used in Figure \ref{fig:cgm_vs_ism_mass}.
        }
        }
        \label{fig:mcCarthy estimates}
    \end{figure}
    \adchange{The restoring pressure due to the gravitational potential of a thin ISM, can be expressed as $2\pi G\Sigma_* \Sigma_{\rm ISM}$, where $\Sigma_{*, \rm ISM}$ is the surface density of the stars and the ISM gas, respectively. However, this estimate for the restoring pressure must be modified in simulations as the modeled disk is not infinitesimally thin.
    We instead consider the maximum restoring pressure on the ISM to be determined by the maximum gravity $g_{\rm max}(R) = {\rm max} (\partial_z \Phi(R))$  along the vertical direction of the disk, such that the Gunn \& Gott criterion is modified to 
    $P_{\rm ram}> \partial_z \Phi(R)_{\rm max} \Sigma_{\rm   ISM}(R)$ (see \citealt{Roediger2005}). The gray curve in Figure \ref{fig:mcCarthy estimates} shows this anchoring pressure for the ISM of our model galaxy. 
    Various horizontal dash-dotted lines represent the ram pressure strength $P_{\rm ram}=\rho v_{\rm rel}^2$ for our simulations and are colored according to the ram pressure strengths shown in Figure \ref{fig:cgm_vs_ism_mass}. Regions of the ISM where the incident ram pressure (dash-dotted lines) exceeds the restoring pressure (gray curve) are expected to get stripped. We find that the mass evolution of ISM seen in the top panel of Figure \ref{fig:cgm_vs_ism_mass} is in accordance with this modified Gunn \& Gott criterion.
    
    \citealt{McCarthy2007} suggest a similar condition for CGM stripping, in which the restoring pressure is expressed as the product of the maximum gravitational acceleration along the vertical direction $\partial_z \Phi(R)_{\rm max}$ and the surface density of gas in the CGM $\Sigma_{\rm CGM}$. The profile of such a restoring pressure for the CGM of our model galaxy is shown in the solid orange line in Figure \ref{fig:mcCarthy estimates}. 
    For the high ram pressures runs, the \citealt{McCarthy2007} criterion suggests a negligible mass of CGM to survive, which is also seen in our simulations (the CGM mass drops below 10\% for the red and magenta lines in Figure \ref{fig:cgm_vs_ism_mass}). For the moderately low ram pressure strengths, the CGM within $20-30\rm \ kpc$ can survive according to  \citealt{McCarthy2007}. The mass loss rate, however, would depend on the ambient density and the relative velocity, as seen in the slope of \textit{thin green} and \textit{dashed green} lines in the bottom panel of Figure \ref{fig:cgm_vs_ism_mass}.
    This is 
    expected at large radii, as both the surface density $\Sigma_{\rm CGM}$ and the maximum gravitational acceleration $\partial_z \Phi(R)_{\rm max}$ fall as $1/R$ (assuming an isothermal potential), implying a restoring pressure $\propto 1/R^2$ as opposed to a constant ram pressure (see Fig. \ref{fig:mcCarthy estimates}). The mass beyond the predicted stripping radius by \citealt{McCarthy2007}, is thus lost at a rate set by the hydrodynamic 
    drag time 
    (see Eq. \ref{eq:mass_loss_rate}). Thus, we agree with the \citealt{McCarthy2007} criterion for the radius outside which the CGM is stripped, but the CGM stripping timescale is given by the appropriate drag time depending on $\chi$.
    }
    
    It is also plausible that CGM stripping starts even before the satellite galaxy enters the parent cluster (see \citealt{Mohammadreza2019, Mohammadreza2021}) -- in the intergalactic medium (IGM). Our simulations of the equilibrium model (\adchange{PG2e-5v800}; see Section \ref{sec:strippingResults}; \adchange{the lowest ram pressure run}) suggest that the CGM and the disk of moderately massive galaxies (like JO201) can survive for long at cluster outskirts. The velocities and densities expected in the IGM are lower than the lowest ram pressure considered in our simulations.
    Given that the CGM survives for several Gyrs in our lowest ram pressure run, the CGM stripping for massive galaxies in the IGM is expected to be \adchange{much less and slower. 
    
    Feedback processes like galactic outflows can push a good fraction of baryons away from the deepest part of the galactic potential well, forming a diffuse, puffed-up atmosphere around the ISM. 
    This weakly bound diffuse hot gas 
    will be stripped easily by the ram pressure like our loosely bound CGM.
    A combination of external ram pressure and internal feedback processes, like supernovae and AGN-driven outflows, is perhaps 
    why a major fraction of infalling satellites in Illustris TNG50 is 
    completely devoid of their CGM (\citealt{Rohr2023}).
    }

\section{Summary}\label{Conclusions}
We have carried out 3D hydrodynamic simulations of model galaxies with both their ISM and CGM, immersed in a constant ICM wind. We have built a simple model of the gas distribution of a galaxy with a rotating ISM disk and a surrounding stationary CGM, which transitions smoothly to the host ICM conditions. Our model galaxies are self-consistent solutions to the steady hydrodynamic equations and \adchange{remain} stable in isolation. Our hydrodynamic simulations suggest that including the CGM in idealized RPS simulations has non-trivial effects. The key results from our study are summarized as follows.
\begin{enumerate}[wide]
    \item 
    We find that once the CGM gains enough momentum, it mediates the ram pressure to the ISM even before the ICM directly comes in contact with the disk. The ISM stripping depends on the ram pressure (\citealt{GunnGott1972ApJ} estimate) while the CGM stripping is akin to \textit{cloud crushing} simulations as gravitational attraction \adchange{on the bulk of the CGM} is negligible.  

    \item Results from our ram pressure stripping (RPS) simulations (see Section \ref{subsec:fiducial_run}) of a galaxy like JO201/JO206 suggest that the CGM would survive for long for galaxies in cluster outskirts. On the contrary, for smaller cluster-centric distances, the CGM is expected to be significantly stripped by $\sim 500$ Myr. This implies that satellites with small impact parameters, in their first pericentric passage, would lose a significant fraction of their CGM. Loss of the CGM would result in a restrained supply of cold gas to sustain star formation, leading to the quenching of star formation. 

    \item We find that the uplifted ISM mixes with both the CGM and the ICM forming long tails of stripped material with coherent rotational and wobbling signatures carried over from the disk. A small fraction of the CGM lingers close to the disk even at $1\ \rm Gyr$ (see Figure \ref{fig:rps_snapshots}). This implies that the metallicity gradients and the rotational signatures observed in the stripped tail would be influenced by the presence of the CGM even when the bulk of it is lost. 
            
    \item In Figure \ref{fig:cgm_vs_ism_mass}, we show that CGM stripping depends on its density contrast (similar to the cloud crushing problem) with the ICM (see Section \ref{subsubsec:cgm_vs_ism}), unlike the disk stripping criterion, which depends on the ram pressure. Comparison of RPS simulations with our control setups without gravity reveals a \adchange{
    quantitatively similar evolution of the} 
    CGM mass. 
    This similarity is studied further in a suite of \textit{`cloud-wind interaction'} simulations of uniform clouds/bubbles immersed in a uniform wind (see Section \ref{subsec:CloudCrushingComparison}).

    \item Given the range of CGM and ICM densities and temperatures, we find that there are two distinct regimes for CGM stripping, depending on its density contrast $\chi$ with the ICM:
    \begin{enumerate}[leftmargin=*, align=left]
    \item the $\chi >1$ regime or the well-known \textit{`cloud crushing'} regime,  where the CGM is denser than the ICM and stripping timescale for the CGM depends on its density contrast with the ICM. In this regime, the CGM mixes with the background ICM due to sustained shear-driven hydrodynamic instabilities. 
    \item the $\chi <1$ regime or the \textit{`bubble drag'} regime, where the CGM is rarer than the ICM and shows a shorter stripping timescale, comparable to the crossing/drag time $t_{\rm drag} \propto R/v_{\rm rel}$ independent of $\chi$ \adchange{and a much reduced mixing}. This regime of \textit{`cloud-wind interaction'} (relevant to CGM-ICM interaction) has been hitherto unexplored in the astrophysical context.
    \end{enumerate}
    
    
\end{enumerate}

\adchange{We find that the} timescale estimates from simplified \textit{`cloud-wind interaction'} problem, \adchange{are in} good agreement with the CGM stripping timescale in our simulations. 
We have explored the \textit{bubble drag} problem in detail and emphasized the need for the \textit{virtual added mass} term 
to explain the evolution of a diffuse bubble (the CGM in our case) moving through a dense wind (the ICM; with high densities at smaller cluster-centric distances).

Overall, our work provides an improved understanding of the hydrodynamic interactions involving ISM-CGM-ICM \adchange{for galaxies falling through cluster/group environments}. This lays the foundation 
on which more complex simulations with radiative cooling, feedback heating, self-gravity, magnetic fields, and turbulence can be built and analyzed.

\section{Acknowledgements}
\adchange{We thank the anonymous referee for insightful comments on the initial manuscript of our paper, which greatly improved the quality of this work.} The authors acknowledge the support from a National Supercomputing Mission Grant and the Supercomputer Education and Research Centre (SERC) at the Indian Institute of Science (IISc). \adchange{RG is grateful to Max Gronke for his valuable inputs throughout the course of this work and to Stephanie Tonnesen for her detailed comments on the manuscript. RG acknowledges} Bianca Poggianti, Antonino Marasco, Simon White, and Ming Sun for stimulating discussions. RG also wishes to thank Dylan Nelson and Rahul Ramesh for discussions on the present understanding of CGM. Research of AD is supported by the Prime Minister's Research Fellowship (PMRF) from the Ministry of Education (MoE), Govt. of India. \adchange{We are grateful to Andrea Mignone and other {\tt PLUTO} developers. We acknowledge the prompt support with in-situ visualization using {\tt ParaView Catalyst} from the members of the \clicks{https://discourse.paraview.org/}{\tt ParaView Discourse} community, especially Jean M. Favre, François Mazen, and Cory Quammen. Several analysis scripts in our work make use of {\tt Numpy} (\citealt{Harris}), {\tt Matplotlib} (\citealt{Hunter}), {\tt Scipy} (\citealt{2020SciPy-NMeth}) and {\tt Cupy} (\citealt{cupy_learningsys2017}) whose communities we thank for
continued development and support.}
\adchange{
\section*{Data Availability}
All the simulation, analysis, and visualization codes used in this work are hosted on the following {\tt GitHub} repository -- \clicks{https://github.com/RitaliG/rps_cgm_hydro}{\tt https://github.com/RitaliG/rps\_cgm\_hydro}. Data related to this work will be shared upon reasonable request to the corresponding author.}


\bibliographystyle{mnras}
\bibliography{rps} 


\appendix


\section{Grid resolution}\label{app:grid}
Our simulations are run in a static grid in 3D polar coordinates $(R, \phi, z)$. Our constructed grid has a uniform high resolution around the disk to resolve the ISM. To prevent the computations from becoming prohibitively expensive, we decrease the resolution gradually in the CGM along $R$ and $z$ directions while the azimuthal resolution is kept fixed. The low resolution at larger galactocentric distances is employed using the `stretched' grid feature of {\tt PLUTO}, for which the cell size grows in geometric progression. The first cell at the start of the stretched grid (say in the radial direction) has a size $\kappa \Delta r$, where $\Delta r$ is the resolution of the nearest uniform grid and $\kappa>1$ is the common ratio of the geometric progression that keeps on increasing the cell size (as $\kappa^2 \Delta r$, $\kappa^3 \Delta r$, ...) till the end of the domain. Similar stretching is used in the vertical ($\hat{z}$) direction as well. Interested readers are referred to the \href{http://plutocode.ph.unito.it/documentation.html}{{\tt PLUTO User's Guide}}.\footnote{\url{http://plutocode.ph.unito.it/documentation.html}} The details on the exact numerical values of cell sizes and the number of grid points used in our simulation are discussed in Section \ref{subsec:gridding}. 
\begin{figure}
    \includegraphics[width=\columnwidth]{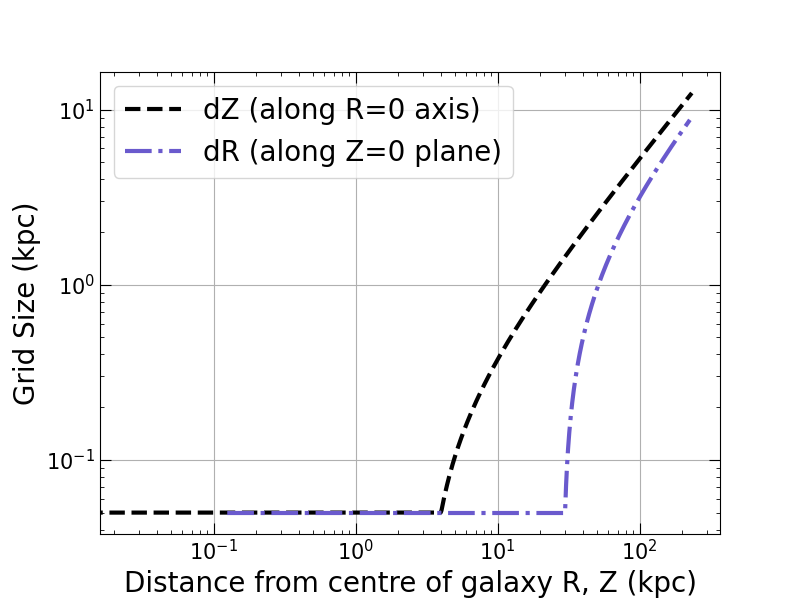}
    \caption{The grid resolution in our simulation domain along the radial (\textit{violet} line) and vertical (\textit{black} line) directions for the fiducial run (OG1e-3v800). The disk has a resolution of $50$ pc, till $R_{0, \rm ISM}$ along $r$ and till $z_{0, \rm ISM}$ along $z$. This progressively becomes stretched by a factor, such that the lowest resolution is around $10$ kpc in the outskirts ($240$ kpc from the disk).}
    \label{fig:stretched grid}
\end{figure}

\bsp	
\label{lastpage}
\end{document}